\documentclass[a4paper,fleqn,5p]{cas-dc}

\usepackage[numbers]{natbib}
\usepackage[english]{babel}
\usepackage{amsmath,amssymb}
\usepackage{graphicx}
\usepackage{array}
\usepackage{multirow}
\usepackage{booktabs}
\usepackage{adjustbox}
\usepackage{float}
\usepackage{xcolor}

\begin{document}


\shorttitle{Privacy-Preserving Architectures for IoT and Vehicular Data Sharing}

\title[mode = title]{Systematic Survey on Privacy-Preserving Architectures for IoT and Vehicular Data Sharing: Techniques, Challenges, and Future Directions}


\author[1,2]{Phat T. Tran-Truong}[type=editor,
                       auid=000,bioid=1,
                       prefix=,
                       role=,
                       orcid=0000-0003-3199-6333]
\fnmark[1]
\ead{phatttt@hcmut.edu.vn}
\credit{Conceptualization, Methodology, Software, Validation, Writing - Original Draft, Writing – review \& editing}
\address[1]{Department of Software Engineering, Faculty of Computer Science and Engineering, Ho Chi Minh City University of Technology (HCMUT), 268 Ly Thuong Kiet Street, Dien Hong Ward, Ho Chi Minh City, Vietnam}
\address[2]{Vietnam National University Ho Chi Minh City, Linh Trung Ward, Ho Chi Minh City, Vietnam}

\author[3]{Vinh X. Q. Nguyen}
\credit{Conceptualization, Methodology, Software, Validation, Writing - Original Draft, Writing – review \& editing}
\address[3]{FPT University, Can Tho, Vietnam}
\ead{vinhnxqce170144@fpt.edu.vn}

\author[4]{Ha X. Son}[orcid=0000-0002-5336-0566]
\ead{ha.son@rmit.edu.vn}
\credit{Conceptualization, Methodology, Software, Validation, Writing - Original Draft}
\address[4]{The Business School (TBS), RMIT University, Ho Chi Minh City, Vietnam}

\author[5,6]{Phien Nguyen-Ngoc}[orcid=0000-0001-9152-6208]
\ead{nguyenngocphien@tdtu.edu.vn}
\credit{Conceptualization, Methodology, Software, Validation, Writing - Original Draft}
\address[5]{Center for Applied Information Technology, Ton Duc Thang University, Ho Chi Minh City, Vietnam}
\address[6]{Faculty of Information Technology, Ton Duc Thang University, Ho Chi Minh City, Vietnam}
\cormark[1]

\author[1]{Khanh H. Vo}
\credit{Methodology, Supervision}

\author[1]{Triet M. Nguyen}
\credit{Methodology, Supervision}

\cortext[cor1]{Corresponding author.}

\fntext[fn1]{First author}

\newcommand{\revisedVersion}[1]{{\color{black}#1}}
\newcommand{\rev}[1]{{\color{black}#1}}

\begin{abstract}
The proliferation of Internet of Things devices and vehicle-to-everything communication systems generates unprecedented volumes of sensitive data at the network edge, necessitating privacy-preserving architectures that enable secure data sharing without exposing raw information to untrusted parties. However, contemporary solutions face a fundamental privacy-efficiency-trust trilemma where achieving strong privacy guarantees, computational efficiency for resource-constrained devices, and decentralized trust establishment simultaneously remains intractable with monolithic single-paradigm approaches. Federated Learning prioritizes efficiency but suffers Byzantine poisoning attacks degrading accuracy from 90\% to 21\%, cryptographic approaches provide provable security but impose prohibitive overhead of 5.7 to 28.4 times longer execution, and blockchain-based solutions establish trust but face scalability bottlenecks with gas costs reaching 1.57 million Gwei for 10 megabyte files.

This survey provides systematic analysis of 75 technical papers published between 2007 and 2025, organized into a novel three-dimensional taxonomy classifying architectures into Decentralized Computation, Cryptography-based approaches, and Distributed Ledger technologies. Our temporal analysis reveals dramatic acceleration during 2024--2025 when 48\% of all papers were published, with Decentralized Computation emerging as the dominant paradigm accounting for 44\% of contributions and 59\% of 2025 publications. Through comprehensive Security Threat Mapping and Technology Maturity Assessment, we demonstrate that mature solutions occupy narrow design space regions excelling in one or two dimensions while compromising others. Our cross-paradigm evaluation conclusively validates the trilemma hypothesis, showing that single-paradigm solutions fundamentally cannot satisfy all three objectives simultaneously.

We identify emerging hybrid architectures combining complementary paradigms—achieving 93.2\% accuracy with 30\% communication reduction—as the essential path forward, articulating critical challenges including security guarantee composition across layers, multi-layer coordination overhead minimization, and post-quantum security integration that must be addressed for practical deployment in next-generation intelligent transportation systems and IoT ecosystems.
\end{abstract}




\maketitle




\section{Introduction}
\label{sec:introduction}

The proliferation of Internet of Things (IoT) devices and vehicle-to-everything (V2X) communication systems has catalyzed unprecedented data generation at the network edge, creating immense opportunities for collaborative learning, predictive analytics, and real-time decision-making. However, the sensitive nature of IoT and vehicular data—encompassing location traces, driving behaviors, and infrastructure vulnerabilities—renders centralized data aggregation increasingly untenable under stringent privacy regulations. Consequently, privacy-preserving architectures that enable secure data sharing without exposing raw data to untrusted parties have become critical enablers for next-generation distributed intelligence systems.

Despite substantial research efforts, contemporary privacy-preserving architectures face a fundamental design constraint that we term the \textit{privacy-efficiency-trust trilemma}. This trilemma posits that achieving strong privacy guarantees, computational efficiency suitable for resource-constrained devices, and decentralized trust simultaneously remains fundamentally intractable with monolithic single-paradigm solutions. \rev{Consider, for instance, a Federated Learning (FL) architecture designed for efficiency and data minimization \cite{lai2022fedscale, ganesh2025data}. By design, it avoids the computational burden of cryptography and the overhead of a global ledger. However, this efficiency comes at the cost of trust. The absence of a decentralized trust layer exposes the system to malicious participants, who can launch Byzantine poisoning attacks that degrade global model accuracy from 90.05\% to as low as 21.35\% on the CIFAR-10 dataset \cite{lin2024sf}. Similarly, cryptographic approaches like Homomorphic Encryption can enforce that only trusted parties participate in training, directly addressing the vulnerability of FL. Yet, this security guarantee imposes a prohibitive computational overhead, executing 5.7 to 28.4 times slower than plaintext operations in common tasks like medical imaging and MNIST classification \cite{mahato2024privacy, jin2025dmafl}.}  Blockchain-based solutions, which introduce a decentralized trust layer to address both malicious participants and single points of failure, face their own critical bottleneck. Traditional Proof-of-Work (PoW) mechanisms, while secure, require block times of over ten minutes, rendering them fundamentally incompatible with the latency-sensitive applications that define V2X communication. Conversely, lightweight blockchain alternatives that promise faster finality often achieve it by redistributing—and effectively centralizing—trust in a small set of validators \cite{alladi2022comprehensive}. \rev{Promisingly, recent innovations in consensus design, such as the intermittent mining protocol Sprints, demonstrate a path to decouple decentralized trust from prohibitive energy costs, paving the way for more sustainable architectures \cite{mirkin2024sprints}. However, the translation and validation of these advances within the unique operational constraints of vehicular data sharing—including sub-second latency requirements and highly dynamic validator sets—remains an open research problem.} 

This trilemma is not a technological impasse but a fundamental property of the design space: optimizing any two objectives necessarily forces a compromise on the third. This inherent tension underscores the need for a holistic understanding of the field, which existing surveys fail to provide. Most focus on individual paradigms in isolation—cryptography-centric surveys omit distributed learning \cite{lu2022survey, daniel2017analysis}, federated learning reviews exclude cryptographic alternatives \cite{ahmad2025enhancing, elzemity2024privacy}, and blockchain surveys treat other paradigms only peripherally \cite{alladi2022comprehensive, platt2023sybil}. Furthermore, many conclude coverage before 2022, missing the most transformative period in the field's recent history (2022–2025) \cite{lu2022survey, yang2021survey}. This fragmented landscape leaves critical questions unanswered for practitioners and researchers alike: When is federated learning preferable to homomorphic encryption? How can complementary paradigms be integrated into hybrid architectures that transcend the limitations of monolithic approaches?

This survey addresses these gaps through systematic analysis of 75 technical papers published \rev{between 2007 and 2025. The year 2007 is selected as the starting point because Bethencourt et al. \cite{bethencourt2007ciphertext} introduced Ciphertext-Policy Attribute-Based Encryption (CP-ABE), establishing the framework for policy-based encrypted access control that underpins all ABE-based IoT schemes analyzed; and He et al. \cite{he2007pda} proposed the first privacy-preserving data aggregation scheme for resource-constrained sensor networks, conceptually prefiguring secure aggregation in modern Federated Learning architectures. Moreover, Liu et al. introduced Private key management in hierarchical identity-based encryption, exploring strategies to prevent key escrow \cite{liu2007private}.} Our principal contributions are fourfold:

\begin{itemize}
    \item \textbf{Comprehensive Taxonomy and Systematic Review:} We propose a three-dimensional taxonomy classifying systems into Decentralized Computation (Federated Learning, Trusted Execution Environments), Cryptography-based approaches (Homomorphic Encryption, Attribute-Based Encryption), and Distributed Ledger technologies (Blockchain, IPFS). Following PRISMA guidelines, we analyze 75 papers selected through rigorous four-stage screening, ensuring reproducibility and minimizing selection bias.

    \item \textbf{Temporal Trend Analysis:} We quantify research evolution from 2007 to 2025, revealing dramatic acceleration during 2024--2025 when 36 papers (48\% of the corpus) were published. Our analysis identifies Decentralized Computation as the dominant paradigm (44\% of contributions), with a striking 2025 surge where Group 1 contributes 10 of 17 papers (59\%), establishing distributed learning and hardware isolation as the current research frontier.

    \item \textbf{Strategic Evaluation Framework:} We provide Security Threat Mapping correlating threats with countermeasures across paradigms, quantifying trade-offs— \allowbreak homomorphic encryption mitigates gradient inference but incurs 5.7--28.4$\times$ overhead while Byzantine-robust aggregation achieves 80\%+ accuracy improvement but requires complex mechanisms \cite{pan2025privacy,mahato2024privacy,jin2025dmafl}. Our Technology Maturity Assessment reveals that production-ready solutions occupy narrow design space regions, excelling in one or two dimensions while compromising others.

    \item \textbf{Trilemma Validation and Hybrid Architecture Vision:} Our cross-paradigm analysis validates the trilemma through empirical evidence: Decentralized Computation achieves efficiency but suffers vulnerabilities (90\% to 21\% accuracy degradation \cite{lin2024sf}); Cryptography provides security but imposes costs (5.7--28.4$\times$ overhead \cite{mahato2024privacy,jin2025dmafl}); We identify emerging hybrid architectures—FL with HE and Blockchain achieving 93.2\% accuracy with 30\% communication reduction \cite{ramesh2025privacy}—as the path forward, articulating critical challenges including security guarantee composition, coordination overhead minimization, and post-quantum integration.
\end{itemize}

The remainder of this survey is organized as follows. Section~\ref{sec:related} positions our work relative to existing surveys. Section~\ref{sec:methodology} describes our PRISMA-guided systematic review methodology. Section~\ref{sec:preliminaries} establishes technical preliminaries and our three-paradigm taxonomy. Sections~\ref{sec:group1}--\ref{sec:group3} analyze Decentralized Computation, Cryptography-based approaches, and Distributed Ledger technologies. Section~\ref{sec:strategic} presents a strategic synthesis through Security Threat Mapping and Maturity Assessment, validates the trilemma, and identifies critical open challenges and the path toward hybrid architectures. Finally, Section~\ref{sec:conclusion} concludes with implications for next-generation systems.

\section{Related Work}
\label{sec:related}

Privacy-preserving data sharing in IoT and vehicular networks has been addressed by numerous surveys, each focusing on specific subsets of techniques or application domains. This section reviews existing surveys and positions our work by highlighting the unique contributions and broader scope of our systematic review.


\subsection{Cryptography-Focused Surveys}

Several surveys have examined cryptographic primitives for data privacy, though predominantly from a technique-centric perspective. A study by Lu et al.~\cite{lu2022survey} provides a comprehensive review of cryptographic techniques for big data security, analyzing approximately 280 papers published between 1982 and 2021. Their taxonomy organizes techniques according to data lifecycle stages—storage (authenticated encryption, integrity auditing), sharing (attribute-based encryption, proxy re-encryption), and computing (homomorphic encryption, secure multi-party computation). While this survey offers deep technical analysis of cryptographic primitives and proposes a reference architecture (Z-CABDS), it focuses primarily on general big data systems with limited attention to decentralized architectures. Federated learning is mentioned only briefly as a subset of multi-party computation, and blockchain is treated as an auxiliary application context rather than a core architectural paradigm. Moreover, the survey's coverage concludes at 2021, missing recent developments in privacy-preserving techniques for IoT and vehicular networks between 2022 and 2025.

A study by Daniel et al.~\cite{daniel2017analysis} focuses specifically on Hierarchical Identity-Based Encryption (HIBE) schemes, providing detailed analysis of approximately 25 protocols spanning from 1984 to 2017 to evaluate their suitability for resource-constrained cloud and IoT environments. Their narrow scope addresses authentication and fine-grained access control but does not encompass the broader spectrum of privacy-preserving architectures such as federated learning, trusted execution environments, or distributed ledger technologies. Additionally, their analysis predates the recent proliferation of vehicular-specific privacy requirements and the integration of blockchain-based access control mechanisms that have gained prominence since 2020.


\subsection{Federated Learning and Distributed Machine Learning Surveys}

The emergence of federated learning as a privacy-preserving paradigm has motivated several dedicated surveys. A systematic literature review by Ahmad et al.~\cite{ahmad2025enhancing} examines model robustness in federated learning, analyzing 70 studies published between 2020 and 2025. Their focus on adversarial robustness, Byzantine resilience, and model aggregation strategies provides valuable insights into FL security challenges. However, the survey does not address integration with other privacy mechanisms such as homomorphic encryption or trusted execution environments, nor does it specifically target IoT or vehicular applications where resource constraints and high mobility introduce additional complexities. The absence of comparative analysis with non-FL architectures (e.g., cryptography-based or blockchain-based solutions) limits its utility for practitioners seeking to select optimal architectures for specific deployment scenarios.

A study by ElZemity and Arief~\cite{elzemity2024privacy} surveys privacy threats and countermeasures in federated learning for IoT, reviewing 49 papers published between 2017 and April 2024. While they provide coverage of FL-specific attacks (model inversion, membership inference, Byzantine attacks) and defenses (differential privacy, secure aggregation), their scope excludes alternative privacy-preserving architectures entirely. The survey does not examine scenarios where federated learning may be unsuitable—such as applications requiring strong cryptographic guarantees for ciphertext computation—nor does it compare FL with homomorphic encryption or blockchain-based solutions. Additionally, vehicular networks, which present unique challenges due to high mobility and intermittent connectivity, receive minimal attention.


\subsection{Trusted Execution Environment Surveys}

Trusted execution environments (TEEs) have been examined primarily through the lens of security vulnerabilities rather than comprehensive architectural analysis. Yang et al.~\cite{yang2021survey} survey software side-channel attacks targeting TEEs, analyzing 105 references from 1992 to 2021. Their focus on attack vectors—cache timing, speculative execution, memory access patterns—provides crucial insights for secure TEE deployment. However, the survey does not address TEE integration with other privacy-preserving techniques (e.g., combining SGX with federated learning or homomorphic encryption), nor does it examine TEE applicability to IoT or vehicular networks where hardware heterogeneity and resource constraints may limit deployment options. Furthermore, the survey concludes in 2021, missing recent advances in TEE countermeasures and emerging TEE architectures such as RISC-V Keystone and AMD SEV-SNP introduced since 2022.

A systematization of knowledge paper by Cerdeira et al.~\cite{cerdeira2020sok} analyzes the security of TrustZone-assisted TEE systems, examining 207 bug reports and vulnerabilities identified between 2013 and 2018 to understand how TEE designs handle concurrent execution and inter-core communication. While their work provides valuable insights into TEE architectural trade-offs, it focuses exclusively on system-level security properties without addressing data sharing protocols, privacy-preserving computation, or application-specific requirements for IoT and vehicular networks.


\subsection{Blockchain and Distributed Ledger Surveys}

Blockchain-based privacy preservation has attracted considerable attention, particularly for vehicular and IoT applications. A comprehensive survey by Alladi et al.~\cite{alladi2022comprehensive} examines blockchain applications for securing vehicular networks, reviewing approximately 75 blockchain-based security schemes published between 2016 and 2021. They classify solutions by functionality (secure communication, privacy preservation, authentication, data management) and consensus mechanism (proof-of-work, proof-of-stake, Byzantine fault tolerance). While the survey offers detailed analysis of blockchain-specific challenges in vehicular environments—scalability, latency, computational overhead—it does not examine non-blockchain privacy architectures. Federated learning, homomorphic encryption, and trusted execution environments are either omitted or mentioned only peripherally, despite their growing adoption in vehicular data sharing applications. Consequently, practitioners lack guidance on when blockchain is preferable to alternative architectures or how to integrate blockchain with complementary privacy techniques.

An extensive systematic review by Platt and McBurney~\cite{platt2023sybil} analyzes blockchain consensus mechanisms and their resistance to Sybil attacks, examining 483 papers published between 2008 and April 2021 across twelve industry verticals including IoT, vehicular networks, healthcare, and smart cities. Their classification framework categorizes consensus mechanisms by Sybil resistance strength (strong, limited, none) and identifies that many IoT-specific mechanisms offer only limited protection due to resource constraints, often relying on permissioned architectures. While this survey provides crucial insights into consensus integrity and leader selection security, it focuses exclusively on availability and consistency properties of distributed ledgers. Privacy-preserving computation techniques—homomorphic encryption, federated learning, secure multi-party computation—are largely absent from their analysis. Moreover, trusted execution environments are discussed only as hardware identifiers for Sybil resistance rather than as privacy enclaves for sensitive data processing.

A study by Alsboui et al.~\cite{alsboui2025toward} surveys IOTA-based distributed ledger solutions for IoT, analyzing 56 papers from 2015 to March 2024. Their focus on directed acyclic graph (DAG) architectures and feeless transactions addresses scalability limitations of traditional blockchain. However, the survey is confined to IOTA-specific implementations and does not compare DAG-based ledgers with other privacy-preserving architectures or evaluate their applicability to high-throughput vehicular scenarios. Additionally, the survey lacks quantitative analysis of performance trade-offs and does not provide implementation guidelines for integrating IOTA with existing IoT platforms.

A review by Chen et al.~\cite{chen2022review} examines blockchain-based privacy protection methods for the Internet of Vehicles, analyzing 41 papers published between 2018 and August 2022. Their taxonomy organizes solutions by privacy goal (identity privacy, location privacy, data privacy) and consensus mechanism. However, the survey focuses narrowly on blockchain and does not consider hybrid architectures combining blockchain with federated learning or homomorphic encryption—approaches increasingly prevalent in recent literature. The absence of performance benchmarks and real-world deployment case studies further limits practical applicability.

A study by Ali et al.~\cite{ali2021ensuring} surveys blockchain and smart contract applications for securing IoT communication, analyzing approximately 85 papers with references spanning from the 2008 Bitcoin whitepaper through 2020. While they provide detailed coverage of smart contract vulnerabilities and auditing techniques, the survey does not address privacy-preserving computation (e.g., executing machine learning models on encrypted data) or compare blockchain-based transparency with confidentiality-focused cryptographic approaches.


\subsection{Hybrid and Cross-Architectural Surveys}

A small number of surveys have attempted to bridge multiple privacy-preserving paradigms. Solis et al.~\cite{solis2024exploring} explore the synergy of fog computing, blockchain, and federated learning for secure IoT systems, reviewing 40 papers published between 2016 and August 2023. They identify complementary roles for each technology—fog computing for low-latency processing, blockchain for tamper-evident logging, and federated learning for privacy-preserving analytics. However, their analysis is limited to three-way integration scenarios and does not encompass homomorphic encryption, trusted execution environments, or attribute-based encryption. Additionally, the survey focuses on IoT broadly defined and does not address vehicular-specific requirements such as high-speed handoff, road-side unit coordination, or V2X communication protocols.

A study by Wakili et al.~\cite{wakili2025privacy} presents a comparative analysis of privacy-preserving methods for IoT, examining 47 papers published between 2018 and November 2023 with focus on federated learning, homomorphic encryption, differential privacy, and secure multi-party computation. Their work provides qualitative comparisons of computational overhead, communication cost, and privacy guarantees. However, the survey does not include trusted execution environments or blockchain, and it lacks systematic coverage of hybrid architectures (e.g., FL with TEE, HE with blockchain). The absence of vehicular-specific analysis further constrains its comprehensiveness.

A survey by Faraji et al.~\cite{faraji2024security} examines security and privacy challenges in 5G-enabled IoT and vehicular networks, analyzing 36 papers published between 2013 and 2024. While they address communication security, authentication, and access control, their coverage of privacy-preserving computation is superficial. Federated learning, homomorphic encryption, and blockchain receive only brief mentions without detailed analysis or classification.



A study by Nkomo et al.~\cite{nkomo2018overlay} surveys overlay virtualized wireless sensor networks for industrial IoT, analyzing approximately 169 papers published between 1998 and 2018, focusing on network architecture and virtualization techniques rather than privacy-preserving data sharing. While they discuss secure communication and access control, cryptographic primitives, federated learning, and blockchain are not within their scope.


\subsection{Positioning This Survey}

The surveys reviewed above exhibit significant gaps in scope, coverage, and architectural comprehensiveness. Most focus on a single privacy-preserving technique (cryptography, federated learning, blockchain, or TEE) without comparing across paradigms or examining hybrid architectures. Domain-specific surveys targeting IoT or vehicular networks typically emphasize one class of solutions, leaving practitioners without guidance on selecting among competing approaches or integrating complementary techniques. Furthermore, many surveys conclude their coverage before 2022, missing the substantial body of recent work on privacy-preserving architectures published between 2022 and 2025.

Our survey addresses these limitations through several distinctive contributions. First, we provide \textit{architectural breadth}: unlike prior surveys that focus on individual techniques, we systematically analyze \textit{three architectural paradigms}—decentralized computation (federated learning, split learning, trusted execution environments), cryptography-based approaches (homomorphic encryption, attribute-based encryption), and distributed ledger technologies (blockchain, IPFS)—encompassing 75 technical papers across six technique categories. Second, we offer \textit{domain specificity}: while acknowledging the broader IoT landscape, we place particular emphasis on vehicular networks, analyzing constraints unique to high-mobility environments (intermittent connectivity, road-side unit coordination, V2X protocols). Third, we ensure \textit{temporal currency}: our systematic review covers papers from 2007 through 2025, capturing the most recent advances in privacy-preserving architectures. Fourth, we employ \textit{systematic methodology}: following PRISMA guidelines, our four-stage selection process (identification, screening, eligibility assessment, inclusion) ensures reproducibility and minimizes selection bias, contrasting with the ad-hoc or unspecified selection methods prevalent in prior surveys. Fifth, we provide \textit{quantitative analysis}: our review includes statistical breakdowns of architectural adoption trends, temporal evolution of technique prevalence, and comparative performance metrics extracted from 75 papers. Finally, we present a \textit{unified taxonomy} that classifies architectures by their core privacy mechanism (data locality, cryptographic indistinguishability, verifiable decentralization), enabling principled comparison across disparate techniques and facilitating architecture selection for specific deployment scenarios.

Table~\ref{tab:related_comparison} summarizes the key characteristics of existing surveys and positions our work within the landscape. The comparison highlights our survey's unique combination of architectural breadth (spanning all three paradigms), domain focus (IoT and vehicular networks), temporal currency (2007–2025), and systematic methodology (PRISMA-compliant selection).


\begin{table*}[!t]
\caption{Comparison of Existing Surveys with our Work}
\label{tab:related_comparison}
\centering
\small
\begin{tabular}{|l|c|c|c|c|c|c|c|}
\hline
\textbf{Survey} & \textbf{Year} & \textbf{Papers} & \textbf{Time Range} & \textbf{FL/SL} & \textbf{TEE} & \textbf{HE/ABE} & \textbf{Blockchain} \\
\hline\hline
Lu et al.~\cite{lu2022survey} & 2022 & $\sim$280 & 1982--2021 & Partial & No & Yes & Partial \\
Daniel et al.~\cite{daniel2017analysis} & 2017 & $\sim$25 & 1984--2017 & No & No & HIBE only & No \\
Ahmad et al.~\cite{ahmad2025enhancing} & 2025 & 70 & 2020--2025 & Yes & No & No & No \\
Yang et al.~\cite{yang2021survey} & 2021 & 105 & 1992--2021 & No & Yes & No & No \\
ElZemity \& Arief~\cite{elzemity2024privacy} & 2024 & 49 & 2017--2024 & Yes & No & No & No \\
Cerdeira et al.~\cite{cerdeira2020sok} & 2020 & 207 (bugs) & 2013--2018 & No & Yes & No & No \\
Alladi et al.~\cite{alladi2022comprehensive} & 2022 & $\sim$75 & 2016--2021 & No & No & Partial & Yes \\
Platt \& McBurney~\cite{platt2023sybil} & 2023 & 483 & 2008--2021 & No & Partial & No & Yes \\
Alsboui et al.~\cite{alsboui2025toward} & 2025 & 56 & 2015--2024 & No & No & No & IOTA only \\
Chen et al.~\cite{chen2022review} & 2022 & 41 & 2018--2022 & No & No & Partial & Yes \\
Ali et al.~\cite{ali2021ensuring} & 2021 & $\sim$85 & 2008--2020 & No & No & No & Yes \\
Solis et al.~\cite{solis2024exploring} & 2024 & 40 & 2016--2023 & Yes & No & No & Yes \\
Wakili et al.~\cite{wakili2025privacy} & 2025 & 47 & 2018--2023 & Yes & No & Yes & No \\
Faraji et al.~\cite{faraji2024security} & 2024 & 36 & 2013--2024 & Partial & Partial & Partial & Partial \\
Nkomo et al.~\cite{nkomo2018overlay} & 2018 & $\sim$169 & 1998--2018 & No & No & No & No \\
\hline
\textbf{This Survey} & \textbf{2025} & \textbf{75} & \textbf{2007--2025} & \textbf{Yes} & \textbf{Yes} & \textbf{Yes} & \textbf{Yes} \\
\hline
\end{tabular}
\end{table*}

In summary, existing surveys provide valuable insights into specific privacy-preserving techniques or application domains but lack the architectural breadth, systematic methodology, and temporal currency necessary to guide practitioners in selecting and deploying privacy-preserving architectures for contemporary IoT and vehicular data sharing scenarios. Our work fills this gap by providing a comprehensive, systematic, and up-to-date analysis spanning all major architectural paradigms, with particular attention to the unique challenges of vehicular networks.

\section{Methodology}
\label{sec:methodology}

This survey employs a systematic literature review methodology to ensure comprehensive coverage, reproducibility, and rigor in selecting and analyzing privacy-preserving architectures for IoT and vehicular data sharing. The methodology consists of two primary components: (1) a structured search strategy and screening process following PRISMA (Preferred Reporting Items for Systematic Reviews and Meta-Analyses) guidelines to ensure transparency and minimize selection bias, and (2) a taxonomy-driven statistical analysis that classifies selected works into distinct technical paradigms based on their architectural contributions. This dual approach enables both breadth—capturing the full landscape of relevant research—and depth—providing detailed technical analysis within each architectural category.

\subsection{Search Strategy and Screening Process}
\label{subsec:search_strategy}

The systematic literature search was conducted using standard bibliometric retrieval tools, including Publish or Perish software, which aggregates scholarly publications from multiple authoritative databases including Google Scholar, IEEE Xplore, and Scopus. These databases were selected for their comprehensive coverage of computer science and engineering literature, with IEEE Xplore providing access to technical publications from the IEEE and IET, Scopus offering broad interdisciplinary coverage with rigorous quality filtering, and Google Scholar ensuring completeness by capturing preprints, conference proceedings, and emerging work not yet indexed in traditional databases.

The search strategy employed a Boolean query combining three conceptual dimensions to precisely target relevant literature while maintaining sufficient breadth to capture diverse technical approaches. The query structure was designed as follows:

\begin{equation*}
\begin{aligned}
      & (\text{``Privacy-preserving''} \vee \text{``Secure''}) \\
\land \quad & (\text{``IoT''} \vee \text{``Vehicular Networks''}) \\
\land \quad & (\text{``Federated Learning''} \vee \text{``Homomorphic Encryption''} \\
      & \quad \vee \text{``Blockchain''})
\end{aligned}
\end{equation*}

The first dimension captures the core privacy objective through synonymous terms encompassing both data confidentiality (privacy-preserving) and broader security properties (secure). The second dimension targets the application domain, focusing on Internet of Things deployments and vehicular network scenarios which share common constraints including resource limitations, dynamic topologies, and distributed data generation. The third dimension specifies key technical mechanisms that represent the three major architectural paradigms identified in preliminary exploration: distributed computation (Federated Learning), cryptographic protection (Homomorphic Encryption), and decentralized trust establishment (Blockchain).
The temporal scope of the search spanned from 2007 to 2025, encompassing foundational works that established core concepts through the most recent developments. 
The search was conducted in January 2025, capturing all literature available through the end of 2024 and early 2025 publications.

\begin{figure}[htbp]
    \centering
    \includegraphics[width=0.95\linewidth]{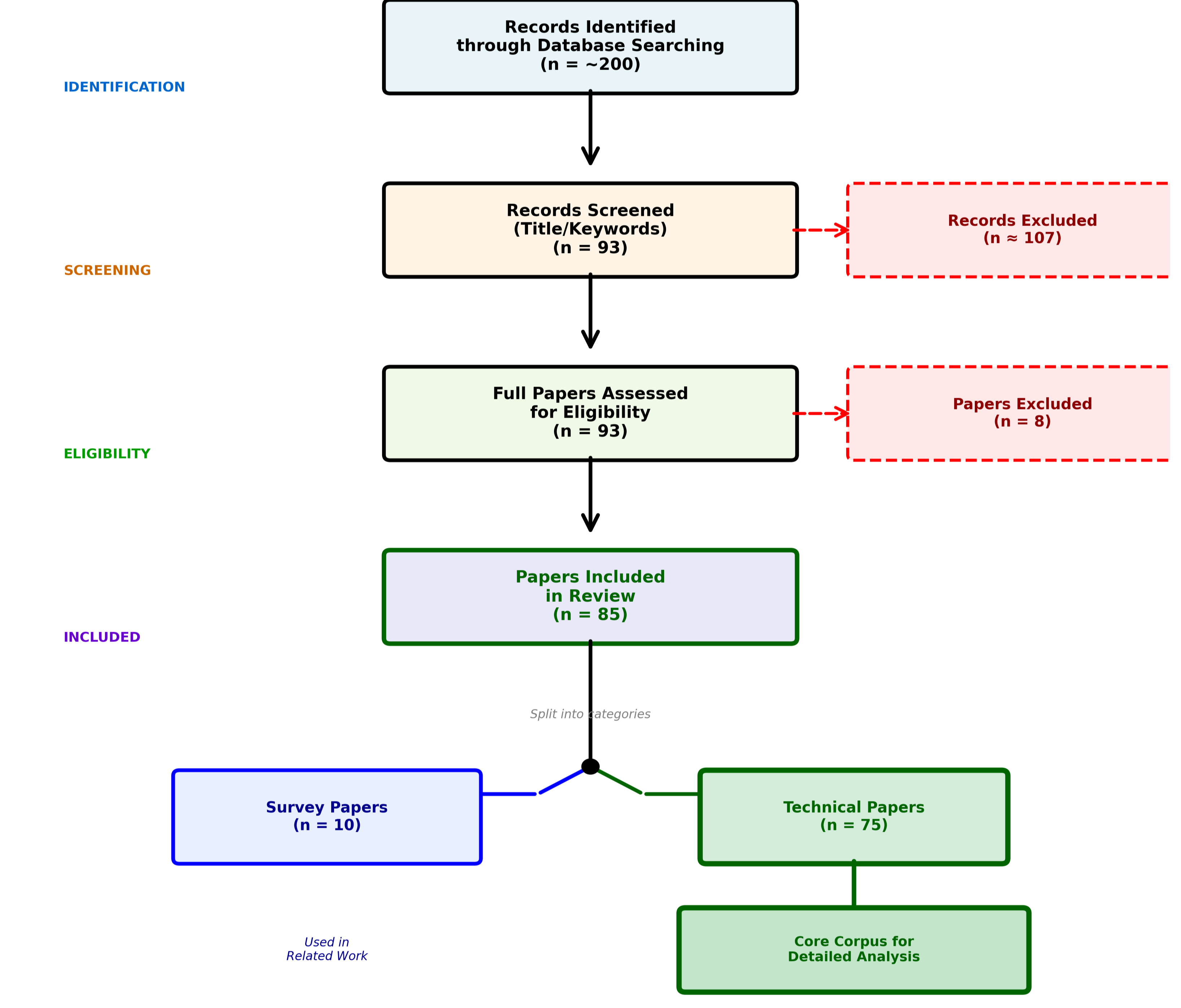}
    
    \caption{PRISMA Flow Diagram illustrating the systematic literature selection process. The methodology follows PRISMA guidelines, searching databases including Google Scholar, IEEE Xplore, and Scopus (2007--2025).}
    \label{fig:prisma_flow}
\end{figure}

The screening process followed a rigorous multi-stage protocol aligned with PRISMA guidelines, as illustrated in Figure~\ref{fig:prisma_flow}. The process proceeded through four sequential stages:

\textbf{Stage 1: Initial Identification.} The Boolean query returned approximately 200 bibliographic records across all queried databases. This initial corpus included journal articles, conference papers, technical reports, and preprints. Automated deduplication was performed to remove exact duplicates resulting from records indexed in multiple databases, though near-duplicates (e.g., conference papers and extended journal versions of the same work) were retained at this stage for manual evaluation.

\textbf{Stage 2: Title and Keyword Screening.} Each record's title and author-provided keywords were manually examined to assess prima facie relevance to the survey scope. Records were excluded if they clearly fell outside the target domain based on the following criteria: (1) papers focusing exclusively on network security (e.g., intrusion detection, authentication protocols) without addressing data privacy during computation or storage; (2) works targeting fundamentally different application domains (e.g., cloud computing, social networks) without applicability to resource-constrained IoT or vehicular networks; (3) purely theoretical cryptography papers without system-level architecture or implementation consideration; (4) duplicate publications where the same work appeared in both conference and journal form, retaining only the most comprehensive version. This screening reduced the corpus to 93 papers deemed potentially relevant based on title and keyword analysis alone.

\textbf{Stage 3: Abstract and Content Analysis.} The abstracts and, when necessary, full text of the 93 papers were thoroughly analyzed to assess substantive contribution and methodological rigor. Papers were evaluated against explicit eligibility criteria: (1) \textit{Technical Contribution}: The work must propose, implement, or rigorously evaluate a privacy-preserving architecture, protocol, or system component rather than merely surveying existing techniques; (2) \textit{IoT/Vehicular Relevance}: The approach must be designed for or evaluated in resource-constrained IoT devices or vehicular network scenarios, or explicitly address constraints characteristic of these domains (limited computation, intermittent connectivity, real-time requirements); (3) \textit{Privacy Focus}: The core contribution must address data privacy, confidentiality, or access control rather than tangential security properties like availability or authentication; (4) \textit{Methodological Soundness}: The work must provide sufficient technical detail to enable understanding of the proposed approach and present empirical evaluation or rigorous security analysis supporting claimed properties.

Eight papers were excluded during this stage for the following reasons: three papers were tangentially related surveys or position papers lacking original technical contributions; two papers focused on orthogonal problems (secure communication channel establishment without privacy-preserving computation); one paper was a duplicate extended version improperly retained during Stage 2; and two papers lacked sufficient technical depth or evaluation rigor to support substantive analysis. This rigorous screening resulted in a final corpus of \textbf{85 papers} deemed highly relevant and methodologically sound.

\textbf{Stage 4: Classification into Technical and Survey Papers.} The 85 papers were further categorized based on publication type and contribution nature. Ten papers were identified as survey, review, or taxonomy papers that synthesize existing literature rather than proposing novel technical solutions. While methodologically sound and valuable for contextualizing the field, these survey papers were separated from the main technical analysis to avoid conflating primary research contributions with secondary literature reviews. These 10 survey papers are instead utilized in the Related Work section to position this survey within the broader landscape of existing reviews and identify gaps that this work addresses.

The remaining \textbf{75 papers} constitute the core corpus for detailed technical analysis. These papers represent original research contributions proposing novel architectures, protocols, algorithms, or implementations for privacy-preserving data sharing in IoT and vehicular networks. The subsequent taxonomy development and comparative analysis focus exclusively on these 75 technical papers to ensure that insights derive from primary research rather than secondary literature.

\subsection{Taxonomy Development and Statistical Analysis}
\label{subsec:taxonomy_analysis}

Following the systematic screening process, the 75 technical papers were subjected to detailed content analysis to develop a taxonomy that organizes the literature based on fundamental architectural paradigms. Unlike superficial categorizations based solely on application domain or deployment setting, our taxonomy classifies works according to their \textit{primary architectural mechanism} for achieving privacy—the core technical approach that distinguishes fundamentally different design philosophies and inherent trade-offs.

\subsubsection{Taxonomy Rationale and Classification Logic}

The taxonomy development process began with exploratory analysis of the 75 papers to identify recurring architectural patterns and technical mechanisms. Three distinct paradigms emerged as the dominant approaches, each addressing the privacy-efficiency-trust trilemma through fundamentally different means:

\textbf{Group 1: Decentralized Computation-based Approaches} leverage distributed learning protocols (Federated Learning, Split Learning) or hardware-assisted isolation (Trusted Execution Environments) to minimize data exposure while maintaining computational efficiency. The core principle is \textit{data locality}—computation is performed where data resides rather than aggregating raw data at central locations. This group encompasses 33 papers (44\% of the corpus), subdivided into:
\begin{itemize}
\item \textit{Federated Learning and Split Learning} (22 papers): Collaborative model training where clients compute local gradients on private data and share only model updates with a central aggregator or peer devices.
\item \textit{Trusted Execution Environments} (11 papers): Hardware-enforced isolation creating secure enclaves where code and data are protected from untrusted operating systems and external adversaries.
\end{itemize}

\textbf{Group 2: Cryptography-based Approaches} employ mathematically rigorous encryption schemes enabling computation on encrypted data (Homomorphic Encryption) or fine-grained access control (Attribute-Based Encryption, Hierarchical Identity-Based Encryption). The core principle is \textit{cryptographic indistinguishability}—encrypted data is computationally indistinguishable from random, preventing information leakage even to parties performing computations. This group comprises 24 papers (32\% of the corpus), divided into:
\begin{itemize}
\item \textit{Homomorphic Encryption} (15 papers): Schemes supporting mathematical operations on ciphertexts that, when decrypted, yield results equivalent to operations on plaintexts.
\item \textit{Attribute-Based and Hierarchical Encryption} (9 papers): Access control mechanisms where decryption capabilities are determined by attribute sets or hierarchical identities rather than individual keys.
\end{itemize}

\textbf{Group 3: Distributed Ledger-based Approaches} utilize blockchain consensus mechanisms and off-chain storage solutions to establish decentralized trust and immutable audit trails without centralized authorities. The core principle is \textit{verifiable decentralization}—security guarantees derive from distributed consensus and cryptographic chaining rather than trust in individual entities. This group includes 18 papers (24\% of the corpus), categorized as:
\begin{itemize}
\item \textit{Blockchain for Access Control and Consensus} (14 papers): Smart contracts and consensus protocols establishing tamper-proof access policies and verifiable data sharing.
\item \textit{IPFS for Off-chain Storage} (4 papers): Content-addressed distributed storage addressing blockchain's storage scalability limitations.
\end{itemize}

Papers were classified into groups based on their \textit{primary architectural contribution}—the core mechanism distinguishing their approach from prior work. Works integrating multiple mechanisms (e.g., Federated Learning with Homomorphic Encryption) were assigned to the group corresponding to their novel technical contribution; if cryptographic enhancement was applied to a federated learning framework, the paper was classified in Group 1 (FL) as the cryptography served as a supporting mechanism rather than the primary architectural innovation. This classification principle ensures that the taxonomy reflects fundamental design paradigms rather than superficial feature combinations.

\subsubsection{Distribution Analysis: Architectural Paradigm Preferences}

\begin{figure}[htbp]
    \centering
    \includegraphics[width=0.8\linewidth]{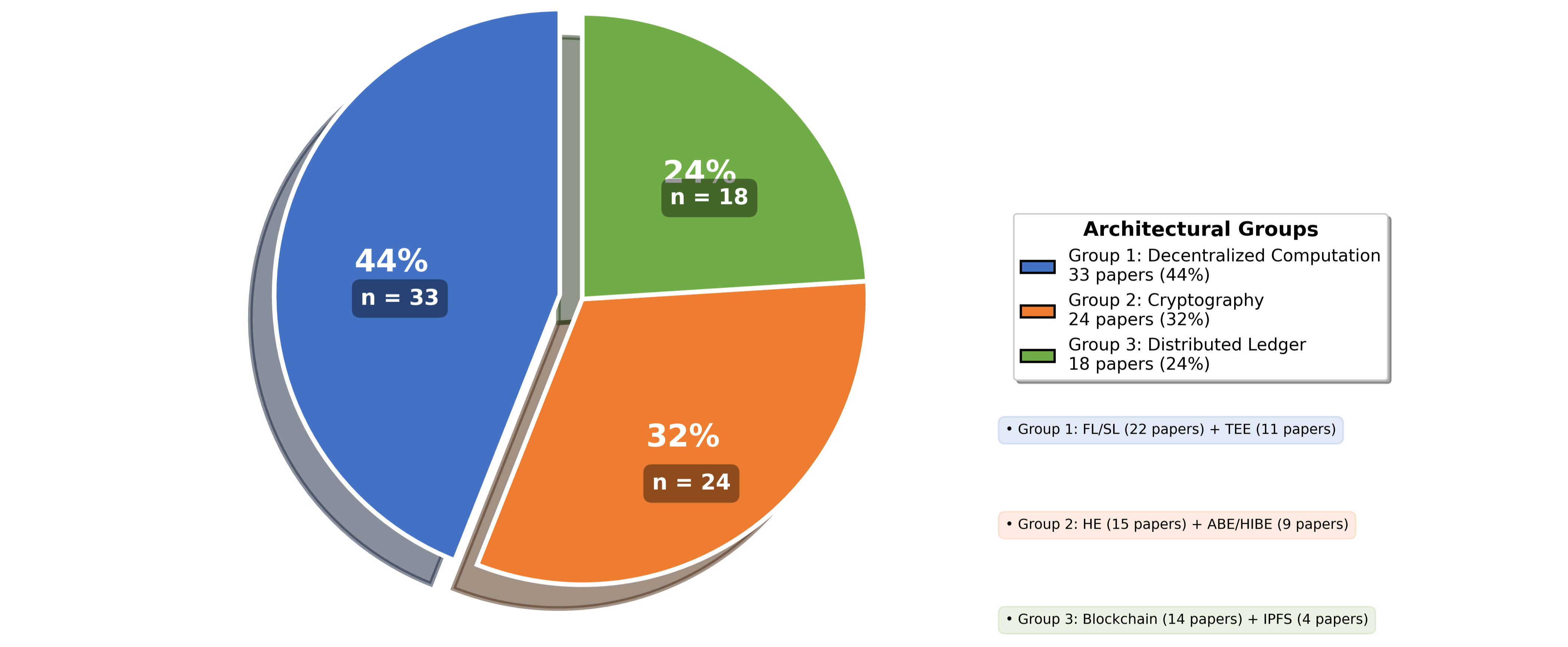}
    
    \caption{Taxonomy distribution of the 75 technical papers (2007--2025).}
    \label{fig:taxonomy_pie}
\end{figure}

The distribution of papers across the three groups, visualized in Figure~\ref{fig:taxonomy_pie}, reveals clear preferences in the research community's architectural approaches to privacy-preserving IoT and vehicular data sharing.

\textbf{Group 1 (Decentralized Computation) emerges as the dominant paradigm, accounting for 44\% of the technical corpus.} This predominance reflects a fundamental shift towards edge intelligence driven by three converging factors. First, the proliferation of computationally capable edge devices (smartphones, modern IoT sensors with embedded processors, vehicular On-Board Units) has made local computation increasingly feasible, eliminating the assumption that raw data must be transmitted to resource-rich cloud servers for processing. Second, Federated Learning has matured from a theoretical framework to a production-ready technology deployed by major technology companies (Google's Gboard, Apple's Siri), providing validated implementations and best practices that lower barriers to adoption. Third, regulatory pressures including GDPR and emerging data localization requirements incentivize architectural approaches that minimize raw data movement across organizational and jurisdictional boundaries, making decentralized computation attractive not merely for technical efficiency but for legal compliance.

The subdivision within Group 1 is particularly revealing: Federated Learning dominates with 22 papers (67\% of Group 1) compared to 11 papers (33\%) on Trusted Execution Environments. This disparity indicates that software-based distributed learning protocols have attracted greater research attention than hardware-assisted isolation, likely due to lower deployment barriers (FL requires no specialized hardware), broader applicability (FL applies to machine learning tasks while TEE protects arbitrary computation), and the current machine learning zeitgeist where privacy-preserving collaborative training addresses timely challenges in model development with sensitive data.

\textbf{Group 2 (Cryptography-based) constitutes 32\% of the corpus, representing a substantial but secondary focus.} Despite the theoretical elegance of cryptographic solutions—particularly Fully Homomorphic Encryption which enables arbitrary computation on encrypted data—practical deployment challenges limit adoption. The high computational overhead of HE (often orders of magnitude slower than plaintext operations) renders real-time applications impractical on resource-constrained IoT devices, confining HE to specific use cases where privacy requirements justify performance penalties. This explains why Group 2, despite offering the strongest mathematical privacy guarantees, attracts less research activity than Group 1's more pragmatic approaches. Within Group 2, Homomorphic Encryption dominates with 15 papers (62.5\%) compared to 9 papers (37.5\%) on Attribute-Based/Hierarchical Encryption, indicating that privacy-preserving computation attracts more attention than access control mechanisms, perhaps because computation represents a harder technical problem requiring cryptographic innovation whereas access control can often be addressed through conventional authenticated encryption and policy enforcement.

\textbf{Group 3 (Distributed Ledger) accounts for 24\% of papers, the smallest group but with concentrated focus on critical infrastructure challenges.} Blockchain's role in privacy-preserving architectures is not privacy provision per se (blockchains are inherently transparent with all data visible to participants) but rather trust establishment and access control enforcement without centralized authorities. The relatively smaller corpus reflects blockchain's supporting role—it provides tamper-proof audit trails and decentralized coordination but must be combined with encryption (Group 2) or local computation (Group 1) to achieve privacy. The subdivision shows blockchain consensus mechanisms (14 papers, 78\%) dominate over IPFS storage solutions (4 papers, 22\%), indicating that establishing lightweight consensus protocols suitable for resource-constrained IoT devices remains the primary challenge, with off-chain storage being a necessary but secondary concern.

\subsubsection{Temporal Trend Analysis: Evolution of Research Focus}

\begin{figure}[htbp]
    \centering
    \includegraphics[width=0.95\linewidth]{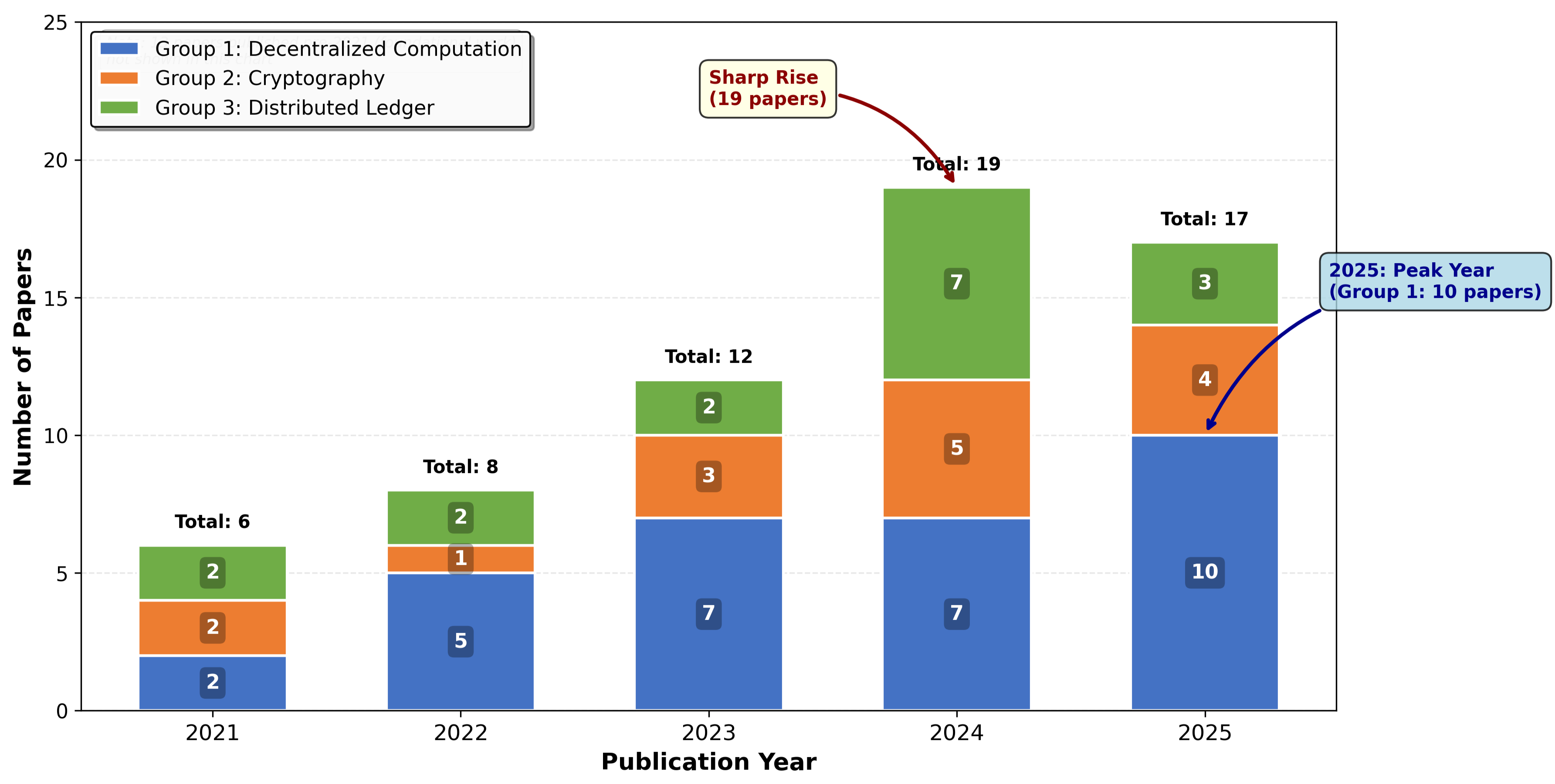}
    
    \caption{Temporal trend of research activity by architectural paradigm (2021--2025).}
    \label{fig:temporal_stacked}
\end{figure}

The temporal distribution of the 75 technical papers, visualized in Figure~\ref{fig:temporal_stacked}, reveals significant insights into the evolution of research priorities and the maturation of different architectural paradigms. To analyze temporal trends while maintaining statistical significance, we aggregate papers published before 2021 (13 papers representing foundational work) and focus detailed analysis on the five-year period from 2021 to 2025 where 62 papers (83\% of the corpus) are concentrated.

\textbf{The most striking observation is the sharp acceleration in research activity during 2024--2025.} Publication volume increases dramatically from 6 papers in 2021 to 19 papers in 2024 and 17 papers in 2025 (through January, suggesting the annual total will exceed 2024 when complete). This surge represents more than simple growth in the field; it indicates a maturation phase where theoretical foundations established in earlier years are being translated into practical systems addressing real-world deployment constraints. The 36 papers published in 2024--2025 alone constitute 48\% of the entire 75-paper corpus, demonstrating that privacy-preserving architectures for IoT and vehicular networks have transitioned from niche research topics to mainstream concerns driving substantial research investment.

\textbf{Group 1 (Decentralized Computation) drives the recent surge, particularly in 2025.} In 2025, Group 1 contributes 10 of 17 papers (59\%), the highest proportion observed in any single year. This dominance suggests that Decentralized Computation has emerged as the current research frontier, with the community converging on distributed learning and hardware isolation as the most promising paths forward. The temporal trajectory within Group 1 is particularly instructive: from 2 papers in 2021 to 7 in 2023, 7 in 2024, and 10 in 2025. This consistent growth, culminating in 2025 as the peak year for Group 1, indicates sustained momentum rather than transient interest. The focus within Group 1 has shifted from foundational Federated Learning frameworks in earlier years towards addressing practical challenges including Byzantine robustness, communication efficiency, and hierarchical aggregation in heterogeneous networks—concerns that must be resolved before large-scale deployment.

\textbf{Group 2 (Cryptography-based) exhibits steady but modest growth.} Publication volume increases from 2 papers in 2021 to 5 papers in 2024, with 4 papers in 2025. This gradual growth without dramatic acceleration suggests that while cryptographic approaches continue to attract research attention, fundamental barriers—particularly computational overhead—limit the rate at which new solutions emerge. The papers in this group increasingly focus on hardware acceleration and optimized implementations rather than entirely new cryptographic schemes, indicating a shift from theoretical cryptography towards systems engineering aimed at making existing schemes practical for resource-constrained devices. The relatively stable year-to-year publication counts suggest that progress in this area is incremental, constrained by the inherent computational complexity of homomorphic operations and pairing-based cryptography.

\textbf{Group 3 (Distributed Ledger) shows the most dramatic recent growth, particularly in 2024.} Blockchain-related papers surge from 2 in 2021--2023 to 7 in 2024, indicating growing recognition that decentralized trust establishment is critical for multi-stakeholder IoT ecosystems. This surge coincides with the maturation of lightweight consensus mechanisms (Proof-of-Authority, Delegated Proof-of-Stake) that make blockchain feasible for resource-constrained devices, overcoming the energy consumption and latency barriers of traditional Proof-of-Work. The modest decline to 3 papers in 2025 (still early in the year) may reflect consolidation as the community converges on established consensus mechanisms rather than proposing fundamentally new approaches. The four IPFS papers are distributed across 2021--2025, indicating sustained but low-volume interest in off-chain storage solutions.

\textbf{Pre-2021 foundational work (13 papers) is heavily dominated by Group 2 (Cryptography).} 
Of the 13 papers published before 2021, 9 papers (69\%) belong to Group 2, reflecting the long-established history of cryptographic research (e.g., HIBE, ABE) compared to the more recent emergence of distributed learning architectures. 
These foundational works established the theoretical frameworks for fine-grained access control that continue to inform current designs. 
In contrast, the minimal presence of Group 1 papers (2 papers) and Group 3 papers (2 papers) in this period underscores that Decentralized Computation and lightweight Blockchain for IoT are genuinely recent phenomena, with the bulk of research emerging only in the last five years.

\textbf{Synthesis: Convergence Towards Hybrid Architectures.} The temporal analysis reveals an important trend that becomes apparent only when examining year-to-year evolution: papers increasingly propose \textit{hybrid architectures} combining mechanisms from multiple groups. While our taxonomy classifies papers based on their primary contribution, many recent works (particularly in 2024--2025) integrate Federated Learning with Homomorphic Encryption, combine Blockchain with IPFS storage, or employ TEE for secure aggregation in FL frameworks. This convergence suggests that the research community increasingly recognizes the limitations of monolithic approaches and pursues architectural integration to transcend the inherent trade-offs of individual paradigms. The sharp growth in Group 1 publications does not imply that decentralized computation alone solves all challenges, but rather that it provides the architectural foundation upon which cryptographic (Group 2) and trust (Group 3) mechanisms are layered to achieve comprehensive privacy-efficiency-trust guarantees.

This temporal trajectory—from scattered foundational work pre-2021, through moderate growth in 2021--2023, to dramatic acceleration in 2024--2025 driven primarily by Group 1—indicates that privacy-preserving architectures for IoT and vehicular networks have reached a critical inflection point. The field is transitioning from exploration of individual techniques to systematic integration of complementary mechanisms, with decentralized computation emerging as the dominant architectural paradigm that other techniques augment. The survey's timing is thus particularly opportune, capturing this maturation phase where sufficient literature exists to identify clear trends, persistent challenges, and promising directions for future research.

\section{Technical Preliminaries}
\label{sec:preliminaries}

This section establishes the theoretical foundation necessary for understanding the privacy-preserving architectures surveyed in this work. We introduce the core cryptographic and computational primitives—Federated Learning, Trusted Execution Environments, Homomorphic Encryption, and Distributed Ledger Technology—that constitute the building blocks of contemporary privacy-preserving systems. Finally, we present the taxonomy diagram that maps these primitives into the architectural paradigms analyzed in this survey.

The privacy-preserving architectures examined in this survey leverage four fundamental classes of technologies, each offering distinct mechanisms for protecting sensitive data while enabling collaborative computation or decentralized coordination. Understanding these primitives is essential for comprehending the architectural trade-offs and design choices analyzed in later sections.

\rev{\subsection{General Architecture for Privacy-Preserving Data Sharing}
\label{subsec:generic_arch}

To clarify the interaction between the entities defined above, we propose a generic architecture illustrated in Fig.~\ref{fig:generic_arch}. This framework demonstrates the multi-layered process required to overcome the privacy-efficiency-trust trilemma \cite{pan2025privacy, lin2024sf}.

As shown in the diagram, the process begins at the \textbf{Data Owner} layer (IoT sensors and vehicles), where sensitive raw data is generated. Instead of direct transmission, the data enters the \textbf{Privacy-Preserving Layer}, which applies specific primitives based on the deployment constraints: (1) Federated Learning ensures data locality by only sharing model updates \cite{han2025privacy}, (2) Homomorphic Encryption provides cryptographic indistinguishability for cloud-based computation \cite{mahato2024privacy}, and (3) TEEs offer hardware-level isolation for security-critical tasks \cite{wang2020trustict}.

The processed data or metadata is then managed by the \textbf{Decentralized Network}, where Blockchain enforces immutable access control policies and IPFS handles scalable off-chain storage \cite{mo2024secure}. Finally, \textbf{Data Consumers} (such as ITS or medical researchers) can retrieve and utilize the information only after successful authentication through the blockchain's smart contracts \cite{ramesh2025privacy}.

\begin{figure}[htbp]
    \centering
    \includegraphics[width=1.0\linewidth]{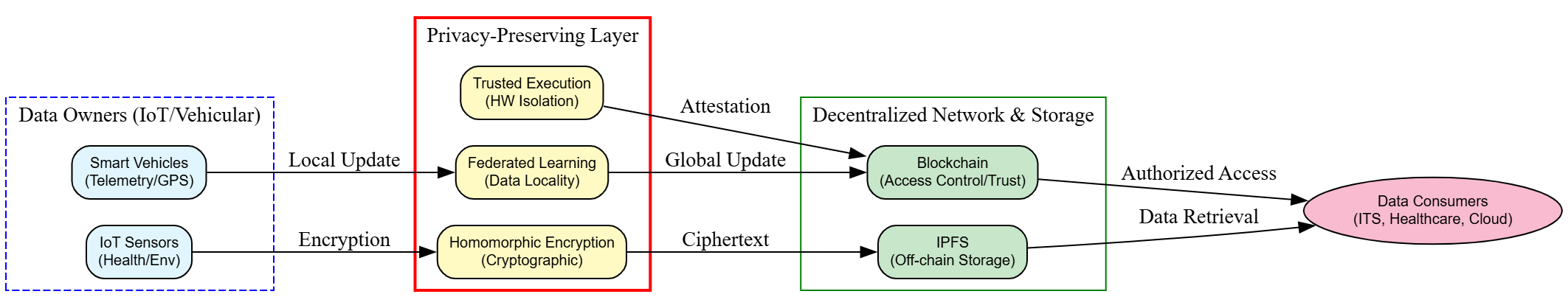}
    \caption{\rev{Proposed generic architecture for privacy-preserving data sharing. It illustrates the data flow from IoT/Vehicular owners through privacy primitives to decentralized storage and authorized consumers.}}
    \label{fig:generic_arch}
\end{figure}}

\subsection{Federated Learning}

Federated Learning (FL) is a distributed machine learning paradigm that enables collaborative model training across multiple participants without centralizing raw data \cite{mcmahan2017communication}. In the canonical FL protocol, clients iteratively compute local model updates (gradients) on their private data and transmit only these aggregated updates to a central server, which combines them into a global model using aggregation algorithms such as Federated Averaging (FedAvg) \cite{mcmahan2017communication}. The fundamental privacy advantage of FL derives from \textit{data locality}: raw training samples never leave individual devices, thereby reducing exposure to potential adversaries compared to traditional centralized learning where all data must be uploaded to a single repository. Variants of FL include vertical federated learning for scenarios where participants hold different feature spaces for overlapping samples \cite{yang2019federated}, and Split Learning where model layers are partitioned between client and server to further minimize communication overhead and computational burden on resource-constrained devices \cite{gupta2018distributed}.

\subsection{Trusted Execution Environments}

Trusted Execution Environments (TEEs) provide hardware-assisted isolation for security-critical code and data, creating protected execution contexts (enclaves) that remain isolated from untrusted software including privileged operating system components and hypervisors. Commercial TEE implementations include Intel Software Guard Extensions (SGX) \cite{costan2016intel, mckeen2013innovative}, which uses processor-enforced memory encryption and access control to create secure enclaves in user-space applications; ARM TrustZone \cite{alves2004trustzone}, which partitions the system into secure and normal worlds with hardware-enforced isolation; and emerging RISC-V extensions for open-source TEE implementations \cite{weiser2019timber}. TEEs enable confidential computing scenarios where sensitive data can be processed on untrusted infrastructure (cloud servers, edge gateways) while maintaining cryptographic guarantees of code integrity and data confidentiality through remote attestation mechanisms that prove to external verifiers that code is executing in a genuine TEE without tampering \cite{coker2011principles}.

\subsection{Homomorphic Encryption}

Homomorphic Encryption (HE) is a class of encryption schemes that support mathematical operations directly on ciphertexts, producing encrypted results that, when decrypted, correspond to operations performed on the underlying plaintexts \cite{rivest1978data}. Partially Homomorphic Encryption (PHE) schemes support either addition (e.g., Paillier cryptosystem \cite{paillier1999public}) or multiplication (e.g., RSA \cite{rivest1978method}), enabling specific computations such as encrypted aggregation or private set intersection. Fully Homomorphic Encryption (FHE) schemes, pioneered by Gentry's breakthrough construction \cite{gentry2009fully}, support arbitrary combinations of additions and multiplications, theoretically enabling any computable function to be evaluated on encrypted data. Contemporary FHE schemes including BGV \cite{brakerski2014leveled}, BFV \cite{fan2012somewhat}, and CKKS \cite{cheon2017homomorphic} employ lattice-based cryptography to achieve practical efficiency for specific applications, though computational overhead remains orders of magnitude higher than plaintext operations. The fundamental advantage of HE is \textit{cryptographic indistinguishability}: ciphertexts leak no information about underlying plaintexts beyond what can be inferred from the computation's output, providing provable security guarantees under well-defined computational hardness assumptions.

\subsection{Distributed Ledger Technology}

Distributed Ledger Technology (DLT) encompasses decentralized data structures and consensus mechanisms that enable multiple mutually distrusting parties to maintain consistent shared state without centralized coordination. \textit{Blockchain}, popularized by Bitcoin \cite{nakamoto2008bitcoin}, uses cryptographic hash chains to create tamper-evident transaction logs where each block contains a hash of the previous block, making historical modifications computationally infeasible without recomputing the entire chain. Consensus mechanisms establish agreement on the canonical blockchain state despite Byzantine adversaries; examples include Proof-of-Work (PoW) requiring computational puzzle solving \cite{nakamoto2008bitcoin}, Proof-of-Stake (PoS) granting voting power proportional to staked cryptocurrency \cite{king2012ppcoin}, and Byzantine Fault Tolerant protocols like Practical Byzantine Fault Tolerance (PBFT) \cite{castro1999practical} achieving consensus through authenticated message passing. Smart contracts—self-executing code deployed on blockchains like Ethereum \cite{wood2014ethereum}—enable programmable access control and automated policy enforcement without trusted intermediaries.

\textit{InterPlanetary File System (IPFS)} addresses blockchain's storage scalability limitations through content-addressed distributed storage \cite{benet2014ipfs}. In IPFS, files are identified by Content Identifiers (CIDs) derived from cryptographic hashes of file contents rather than location-based addresses, enabling decentralized retrieval from any peer hosting the content. IPFS employs Distributed Hash Tables (DHTs) for peer discovery \cite{maymounkov2002kademlia} and incentive mechanisms like Filecoin \cite{protocol2017filecoin} to encourage long-term storage. When integrated with blockchain, IPFS provides scalable off-chain storage while blockchains record CIDs and access policies on-chain, creating hybrid architectures that balance decentralized trust with storage efficiency.

\begin{figure}[htbp]
    \centering
    \includegraphics[width=0.95\linewidth]{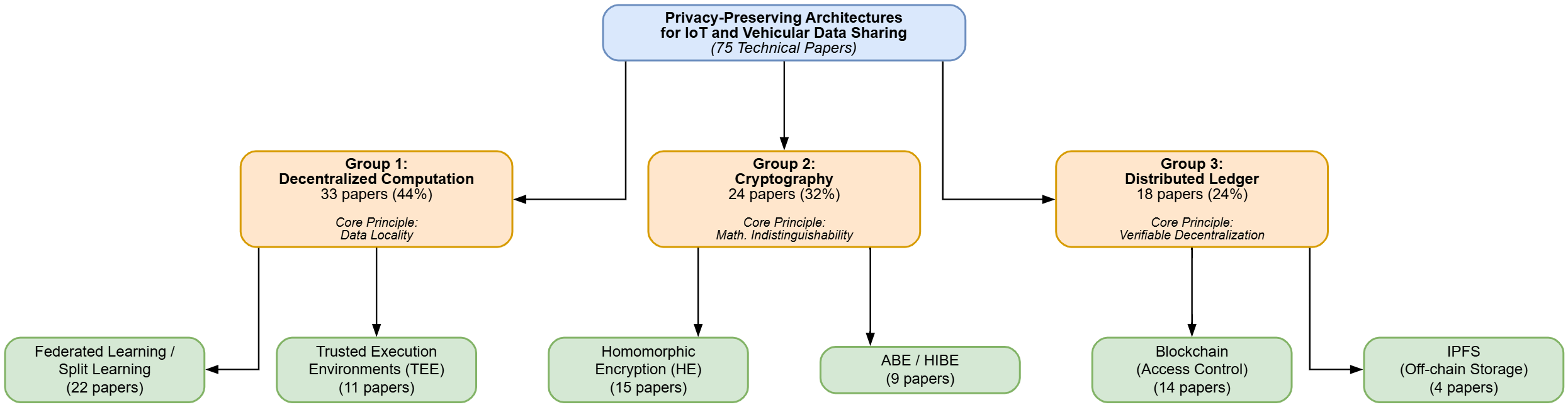}
    \caption{Proposed Taxonomy of Privacy-Preserving Architectures: Three architectural paradigms (Decentralized Computation, Cryptography, Distributed Ledger).}
    \label{fig:taxonomy_tree}
\end{figure}

Based on the fundamental primitives defined above, we structure our analysis using the taxonomy illustrated in Fig.~\ref{fig:taxonomy_tree}.
This classification framework maps the surveyed works into three primary architectural groups, providing a roadmap for the detailed technical analysis in the subsequent sections.

\section{Decentralized Computation-based Approaches}
\label{sec:group1}

Decentralized computation paradigms fundamentally redefine the privacy-efficiency trade-off by eliminating centralized data aggregation. Unlike traditional centralized machine learning, which exposes raw data to a single point of failure, these approaches distribute computational tasks across edge devices (Federated Learning) or enforce hardware-level isolation (Trusted Execution Environments). However, both subgroups face distinct vulnerabilities: FL systems remain susceptible to adversarial model manipulation through Byzantine poisoning and gradient inference attacks, while TEE implementations suffer from microarchitectural side-channel leakage and limited secure memory capacity that constrains the complexity of protected operations. \rev{To address these vulnerabilities, Fang et al. propose a decentralized aggregation framework that eliminates the need for a central trusted server. This approach utilizes Byzantine-robust gradients to maintain model convergence even when a significant portion of the network is compromised, which is particularly relevant for the uncontrolled environments of vehicular networks \cite{fang2024byzantine}.}

\subsection{Federated Learning and Split Learning}
\label{subsec:fl}

Federated Learning has emerged as the predominant privacy-preserving paradigm for collaborative model training in IoT ecosystems, enabling local model updates without raw data transmission to central servers. The fundamental premise—that local gradient updates reveal less information than raw data—has been challenged by recent inference attacks that reconstruct training samples from shared gradients. \rev{Furthermore, comprehensively auditing the exact privacy leakage in these decentralized environments remains challenging. Change et al. \cite{chang2024efficient} quantify membership inference risks by analyzing the trajectory of model performance metrics across FL rounds, effectively evaluating privacy vulnerabilities without incurring the prohibitive computational overhead of training auxiliary attack models, making it exceptionally well-suited for resource-constrained IoT nodes.
By strategically applying gradient compression and noise injection based on eigenspace analysis, this approach neutralizes Data Reconstruction Attacks (DRA) while preserving model utility, offering a scalable and mathematically rigorous defense without the latency of heavy cryptographic operations \cite{tan2024defending}. Beyond inference risks, FL architectures are highly susceptible to sophisticated backdoor and poisoning attacks. To proactively detect such malicious model manipulations, the BackdoorIndicator framework leverages Out-Of-Distribution (OOD) data at the aggregation server. By dynamically injecting synthetic triggers to amplify latent backdoor behaviors, it effectively isolates poisoned local updates even when adversaries attempt to mask their statistical footprints \cite{li2024backdoorindicator}.} \rev{However, providing formal privacy guarantees often degrades utility. The PrivateFL framework addresses this by employing personalized data transformations before applying differential privacy \cite{yang2023privatefl}. This method achieves high accuracy while ensuring that the data of individual IoT nodes remains mathematically indistinguishable from the global aggregate.} Consequently, contemporary FL frameworks increasingly integrate cryptographic primitives to provide provable privacy guarantees.

\subsubsection{Cryptographic Enhancement of Federated Learning}

To mitigate gradient-based inference risks, several works integrate Homomorphic Encryption with FL to ensure that model aggregation occurs on encrypted gradients. The PPVFL scheme employs HE to encrypt local updates before aggregation, achieving a Dice coefficient of 0.918 on medical imaging datasets while incurring an increase in execution time (e.g., from $1.12$s to $1.98$s for 60 clients) due to consensus building in blockchain-based verification systems \cite{mahato2024privacy}. Similarly, threshold CKKS encryption with Shamir's secret sharing maintains accuracy comparable to plaintext Hierarchical FL in both IID and non-IID settings while limiting encrypted model update size to 88 KB, demonstrating collusion resistance against $(n-1)/2$ malicious participants under a semi-honest adversarial model \cite{alzahrani5269991privhfl}. However, encryption introduces predictable runtime increases that scale linearly with the number of participating clients, posing challenges for large-scale deployments.

Multi-layered cryptographic frameworks combine multiple primitives to achieve defense-in-depth. The FL-DABE-BC architecture integrates Decentralized Attribute-Based Encryption for data encryption, HE for secure aggregation, and Secure Multiparty Computation among fog nodes to jointly compute aggregate weights while keeping individual inputs private \cite{narkedimilli2025fl}. Differential Privacy is applied during aggregation to prevent individual data point leakage through statistical analysis of aggregate outputs. While no specific numerical metrics are provided in the available analysis, the framework emphasizes architectural scalability through microservices deployment at the fog layer, with planned integration of Layer-2 solutions like zero-knowledge Machine Learning (zkML) for large-scale computational offloading.

The DP-SA framework integrates Differential Privacy and Secure Aggregation, achieving $96.38\%$ privacy preservation and $95.12\%$ model accuracy \cite{11032530}. The dual-layer security approach—DP for noise injection to obscure individual contributions and SA for encrypted gradient uploads—significantly reduces computational and communication overhead compared to traditional FL methods that employ only a single protection mechanism. The framework demonstrates that carefully calibrated noise addition can preserve utility while providing rigorous privacy guarantees under differential privacy's $(\epsilon, \delta)$-framework.

Advanced schemes leverage post-quantum cryptography to defend against future quantum computing threats. The NTRU lattice-based encryption framework achieves computational efficiency approximately $13\times$ faster than RSA and $5\times$ faster than ECDH-based schemes, with computational complexity of $O(N \log N)$ for key exchange and encryption operations \cite{hota2025advanced}. The scheme maintains $95\%$ accuracy on MNIST and $90.63\%$ accuracy with F1 score of $90.84\%$ for COVID-19 detection tasks. Device computation time decreases as the number of aggregation servers increases—from 125 minutes with 1 server to 100 minutes with 4 servers for parameter $N=503$—demonstrating the benefits of distributed aggregation architectures. The Relay-Agent framework offloads decryption and re-encryption tasks from resource-constrained clients, with gradients sliced into shares to simplify computation and reduce communication overhead.

Efficient aggregation strategies further reduce overhead through gradient compression and selective transmission. The PFed-HE system integrates CKKS encryption, Differential Privacy, and Secure Multiparty Computation, achieving $91$-$94\%$ accuracy with rapid convergence through dynamic key management featuring periodic key rotation and secure key agreement protocols \cite{gadiwala2024enabling}. The framework utilizes batch processing and parallelized HE operations to reduce computational overhead, outperforming reference implementations from Google and Intel in communication efficiency, though specific numerical comparisons are not provided in the available analysis. DMAFL employs Threshold Fully Homomorphic Encryption based on CKKS integrated with Shamir's secret sharing, reducing encryption and decryption overhead by up to $28.4\%$ compared to threshold Paillier \cite{jin2025dmafl}. Under a $40\%$ label-flipping attack, DMAFL achieves only $2.32\%$ accuracy drop on MNIST and $7.44\%$ on CIFAR-10, compared to $7.26\%$ and $10.63\%$ for threshold Paillier respectively. Key generation time is significantly faster, requiring only $13.9$-$90.9\%$ of the time consumed by threshold Paillier across varying key lengths, demonstrating substantial efficiency improvements while maintaining security guarantees.

The framework proposed by Shengxing et al.\ utilizes Top-K gradient selection, candidate quantization protocol, and secure candidate index merging to reduce communication overhead and accelerate HE calculations \cite{shengxing2023efficient}. Gradient clipping and quantization leverage Gaussian distribution characteristics of model parameters to speed up HE computation. An unsigned quantization protocol specifically designed for homomorphically encrypted data further reduces computational complexity. While the paper reports high accuracy and efficient performance, specific numerical results are not provided in the available excerpts.

Tensor decomposition offers an orthogonal optimization approach to reduce model complexity. TT-HeteroFL leverages Tensor Train decomposition to reduce training parameters by $98.81\%$—from 3,277,312 to 38,922 parameters for convolutional layers \cite{zhao2023tensor}. On MNIST, the framework achieves $99.54\%$ accuracy while reducing parameters by $95.14\%$ compared to standard HeteroFL, with encrypted communication size reduced from 393.47 MB to 19.14 MB per round. On CIFAR-10, communication parameters are reduced by $71.13\%$ while maintaining accuracy comparable to full-parameter models. The system employs the BCP cryptosystem (a variant of Paillier) with double trapdoor decryption, where encrypted local models are blinded with random non-zero integers by the FL Platform before transmission to the Cloud Server for secure aggregation.

Healthcare applications demonstrate domain-specific optimizations that balance privacy requirements with clinical utility. The framework integrating FL, Paillier HE, k-anonymity, and Differential Privacy achieves $93.2\%$ accuracy on the MIMIC-III dataset—outperforming both traditional centralized learning ($85.4\%$) and vanilla federated learning ($85.5\%$) approaches \cite{ramesh2025privacy}. Blockchain integration provides a tamper-proof ledger for audit trails, while smart contracts automate data-sharing agreements and enforce access control policies. The framework achieves a $30\%$ reduction in communication overhead compared to traditional FL through selective model update transmission and demonstrates faster convergence through adaptive learning rate scheduling. However, the authors acknowledge that Paillier HE imposes substantial computational overhead, requiring high processing power and longer execution times that may limit deployment on resource-constrained medical IoT devices.

Secure data fusion across heterogeneous IoT frameworks is explored through four complementary algorithms: Secure Aggregation using HE, Unified Learning with Secure Model Aggregation (characteristic of FL architectures), Privacy-Preserving Ensemble Learning, and Secure Multiparty Computation \cite{goyal2024secure}. All four algorithms achieve ``High'' scores for both confidentiality and integrity in security evaluation. Mean Squared Error values range from 0.018 to 0.027 across different algorithms and datasets. Efficiency analysis reveals that Secure Aggregation incurs Low/Low overhead and costs, FL demonstrates Moderate/Moderate resource consumption, Privacy-Preserving Ensemble Learning shows Moderate/Moderate efficiency, while SMC exhibits High/High computational and communication costs due to its interactive nature requiring multiple rounds of communication among parties.

\subsubsection{Byzantine Robustness and Adversarial Defense}

Byzantine attacks represent a fundamental threat to FL systems, where malicious participants inject poisoned updates designed to degrade global model accuracy or implant backdoors that activate under specific trigger conditions. Traditional aggregation schemes like FedAvg are defenseless against coordinated attacks where multiple adversaries collude to overwhelm honest participants.

DRL-PBFL employs Deep Deterministic Policy Gradient to learn a performance-based weighting aggregation policy, allocating lower weights to Byzantine vehicles based on global model performance evaluated on a standard root dataset \cite{pan2025privacy}. The scheme significantly improves global model accuracy compared to existing secure Byzantine FL schemes: by more than $80\%$ on MNIST, more than $30\%$ on CIFAR10, and approximately $15\%$ on Fashion-MNIST under tailored Byzantine attacks designed to maximize damage. The scheme utilizes a secure aggregation algorithm based on Lagrange interpolation, generating eliminable perturbations (masks) that are added to local updates. The cryptographic foundation leverages Shamir's secret sharing (providing additive homomorphism) and naive multiplicative secret sharing (providing multiplicative homomorphism) to enable secure computation without exposing individual model updates.

SF-CABD employs CKKS Homomorphic Encryption with a dual-server architecture to detect Byzantine adversaries while preserving privacy \cite{lin2024sf}. Server 1 handles encrypted data aggregation and preliminary Byzantine detection, while Server 2 (possessing the private decryption key) performs decryption for clustering and anomaly identification. The framework defends against encrypted model poisoning attacks where the number of malicious participants $n^* < n/2$, utilizing clustering based on encrypted Euclidean distance and secure cosine similarity computation to identify malicious gradient vectors that deviate significantly from the honest majority. Under a Bitflip attack on CIFAR-10, SF-CABD achieves $90.05\%$ accuracy compared to $21.35\%$ for FedAvg, demonstrating the severity of the threat and the effectiveness of the defense. The accuracy difference between encrypted (CKKS) and non-encrypted models is negligibly small (approximately $0.05\%$) even with 30 clients, indicating that the cryptographic overhead does not significantly degrade model utility. Batch processing for encrypted Euclidean distance calculation significantly improves encryption efficiency by reducing the number of ciphertext operations.

Ensemble Federated Learning presents an alternative architectural approach to Byzantine defense. FLPhish uses a reputation-based phishing method where clients train locally and send back predictions (not gradients) of an unlabeled public dataset \cite{li2023defending}. The ``phishing'' mechanism uses a labeled dataset (bait $B_t$) maintained by the server to detect attackers by evaluating prediction accuracy—clients whose predictions significantly deviate from ground truth are assigned low reputation scores. A Bayesian inference-based reputation mechanism identifies low-reputation clients as Byzantine attackers and excludes them from aggregation. Under untargeted attacks, FLPhish with reputation-weighted aggregation maintains stable performance even when Byzantine attackers reach or exceed $50\%$ of participants. The Ensemble FL architecture inherently defends against backdoor attacks because model aggregation is based on predicted labels rather than gradient updates, preventing the transfer of backdoor trigger patterns embedded in model parameters. Additionally, EFL significantly reduces communication overhead compared to typical FL by transferring predicted data (typically much smaller than full model parameters) rather than complete model updates.

\subsubsection{Hierarchical and Large-Scale Federated Learning}

Hierarchical Federated Learning addresses communication bottlenecks in flat FL architectures by introducing intermediate aggregation at edge servers, creating a multi-tier structure that reduces long-distance communication between edge devices and cloud servers.

The HFCC framework employs dynamic horizontal splitting to partition Large Language Models into global shared layers and device-local layers, isolating sensitive data processing to the local layer that never leaves the device \cite{han2025privacy}. The framework achieves $58\%$ faster inference speed in resource-constrained environments compared to traditional FL, $94\%$ accuracy in malicious node detection through Byzantine-resistant cross-device consensus verification, $35\%$ reduction in energy consumption, and $42\%$ reduction in communication overhead. The Byzantine-resistant mechanism validates distributed intermediate inference results through voting protocols, with Practical Byzantine Fault Tolerance (PBFT) consensus triggered if disagreement rates exceed predefined thresholds. Overall system analysis reveals a net gain of $7.7\%$ in communication savings and $19.0\%$ in computation reduction when accounting for all architectural components.

Resource allocation strategies for HFL optimize device selection to maximize training efficiency. The device selection strategy based on Deep Deterministic Policy Gradient algorithm considers device data quality, computational capabilities, and communication capacities to identify optimal participants for each training round \cite{yuan2024resource}. On the FASHION-MNIST dataset, the scheme improves model training accuracy by $14.2\%$ compared to DDPG without device selection and by $4.33\%$ compared to DQN with device selection. Training loss is reduced by $45.71\%$ compared to DDPG without selection and by $17.4\%$ compared to DQN with selection. The proposed strategy achieves shorter FL execution times, with efficiency improvements becoming more pronounced as the number of communication rounds or participating devices increases.

\subsubsection{Communication Efficiency Optimizations}

Communication overhead remains the primary bottleneck in resource-constrained IoT deployments, often exceeding computational costs by orders of magnitude due to limited bandwidth and high energy consumption of wireless transmission.

ASAFL employs an adaptive scheduling strategy where clients upload parameters only when their contribution (measured by model distance $\mathcal{D}$) surpasses a dynamic threshold ($\Delta_k^t$), filtering out redundant information that provides minimal improvement to the global model \cite{zhang2025adaptive}. The framework significantly reduces communication overhead by over $70\%$ compared to traditional asynchronous FL (FedAsyn) and by approximately $30\%$ compared to semi-asynchronous approaches (SemiAsyn and SemiSyn). ASAFL demonstrates remarkable convergence performance, often converging by the 20th round in the SVHN/LeNet scenario and achieving up to $98.3\%$ accuracy comparable to synchronous FedAvg. However, ASAFL struggles to counteract severe data heterogeneity, with convergence difficulties observed when the Dirichlet distribution coefficient $\alpha < 0.4$, indicating limitations in highly non-IID settings where local data distributions differ drastically across clients.

Selective Updates and Adaptive Masking are explored through two complementary methods \cite{herzog2024selective}. Selective Updates reduce communication frequency by training until a dynamic threshold on model performance improvement is surpassed, ensuring that only meaningful updates are transmitted. Adaptive Masking performs parameter-level filtering, selecting only model parameters that have changed beyond a threshold ($\delta_n \geq \alpha_k$) relative to the previous round. The combined SU+AM method reduces overall communication volume by over $20\%$ while maintaining model accuracy in standard scenarios. In simulations with 10 FL clients under lower throughput constraints ($33\%$ and $50\%$ of full model size), SU+AM reduces overall bits transferred by up to $37\%$ on CIFAR-10 and up to $29\%$ on MNIST compared to benchmark methods employing deterministic sparsification and stochastic masking (DS+SM).

Layer freezing strategies offer an orthogonal optimization approach. FedGLF gradually reduces the model portion transmitted between server and clients by freezing layers incrementally from input to output layers after they converge to stable values \cite{malan2022communication}. Communication volume savings range from $14.1\%$ to $59.0\%$ for achieving the same target accuracy, depending on the accuracy constraint and data distribution characteristics. In non-IID settings—which are prevalent in real-world IoT deployments—savings reach up to $59.0\%$ for CIFAR-10 and $54.8\%$ for CIFAR-100. Furthermore, FedGLF achieves up to $+2.5\%$ higher accuracy compared to FedAvg while consuming comparable or lower communication costs, demonstrating that the layer freezing strategy does not compromise model quality.

Semi-asynchronous strategies balance the trade-off between model staleness (caused by slow stragglers) and synchronization overhead. FedMDS utilizes spectral clustering to group clients based on both delay characteristics and the direction of model updates, with a synchronous trigger mechanism based on Earth Mover's Distance to limit model staleness \cite{zhang2023fedmds}. Experimental results show FedMDS significantly improves average test accuracy by more than $+9.2\%$ on four federated datasets compared to time-based asynchronous FedAvg (TA-FedAvg), with a maximum accuracy improvement of $+37.6\%$ on the FEMNIST dataset. FedMDS significantly reduces Average Synchronization Waiting Time and alleviates volatility compared with threshold-based FedAvg (T-FedAvg) and group-based FedAvg (G-FedAvg), demonstrating better adaptability to heterogeneous client environments.

\subsubsection{Decentralized and Peer-to-Peer Federated Learning}

Decentralized Federated Learning eliminates the central aggregation server, operating in a peer-to-peer manner to enhance privacy (by avoiding a single point of data collection) and improve fault tolerance (by eliminating the central server as a single point of failure).

DeProFL propagates small prototype representations— \allowbreak vectors computed by the encoder layer representing class centroids—with neighboring devices instead of transmitting large complete model parameters \cite{li2023prototype}. The framework achieves more than $89\%$ accuracy even when the number of samples per class is severely reduced to approximately 10, demonstrating robustness to extreme data scarcity. Compared to the decentralized baseline DeceFL algorithm, DeProFL shows superior accuracy, ranging between $7.7\%$ and $23\%$ higher across different datasets and network topologies. Communication cost per device per round ranges from 0.3 to 0.6 (in normalized units), substantially smaller than algorithms transmitting full model parameters which typically require 10-100 times more communication. However, computational time cost per round is relatively high because each device operates simultaneously as both client and server, performing both local training and aggregation from neighbors.

\subsubsection{Blockchain-Integrated Federated Learning}

The integration of blockchain with FL addresses the trust deficit in federated aggregation by providing tamper-proof audit trails and verifiable computation, enabling detection of malicious aggregation or model poisoning attempts.

The SBSF architecture uses Feldman Verifiable Secret Sharing combined with blockchain to ensure secure storage and privacy preservation of transmitted models \cite{shen2022network}. For convergent models that are stored off-chain, files are encrypted and stored in IPFS in ciphertext format rather than plaintext, with content identifiers recorded on-chain. The framework defends against collusion among up to $(n-1)/2$ malicious participants under the semi-honest adversarial model. Using IPFS for off-chain storage significantly reduces uploaded data size; without IPFS, on-chain data size is approximately $3816\times$ larger. An optimized Raft consensus algorithm based on Named-Data Networking (NDN) improves blockchain network transmission efficiency, with consensus latency decreased by up to $5\%$ and Raft throughput initially higher by $43\%$. However, the total running time of FL with Feldman VSS is nearly $5.7\times$ that of original FL due to the time-security trade-off inherent in cryptographic operations. Feldman VSS encryption time is almost $5.8\times$ the decryption time, indicating asymmetric computational costs.

\subsection{Trusted Execution Environments}
\label{subsec:tee}

Trusted Execution Environments provide hardware-assisted isolation for security-critical operations, creating secure enclaves where code and data are protected from untrusted operating systems, privileged software, and external adversaries through processor-enforced access control. However, TEE implementations face two critical challenges: microarchitectural side-channel vulnerabilities that leak sensitive information through shared hardware state (caches, page tables, branch predictors), and limited secure memory capacity that constrains the complexity and size of protected operations.

\subsubsection{RISC-V Architectures for Lightweight TEE}

RISC-V processors offer a flexible open-source Instruction Set Architecture that enables implementing lightweight TEE solutions tailored to resource-constrained IoT devices without the overhead of complex privilege modes.

A lightweight TEE construction for embedded systems with RISC-V CPUs featuring only Machine (M) and User (U) modes leverages the Physical Memory Protection (PMP) module to provide hardware-based isolation \cite{ma2021construction}. The scheme modifies the standard execution structure so that both the trusted OS kernel and embedded RTOS kernel run in M mode in an isolated manner. The PMP extension introduces access restriction of M mode to Supervisor/User mode memory regions and enables operation isolation within M mode itself, distinguishing between Primary Context (secure monitor, bootloader) and Secondary Contexts (Trusted OS kernel, RTOS kernel). The design improves performance through a zero-copy data mechanism for data transfer in shared memory regions, eliminating context switching overhead for common executable codes that can be shared across security domains.

Secure services for standard RISC-V architectures implement TEE functionality using PMP to manage memory access permissions (read, write, execute) for lower-privileged S-Mode and U-Mode software, partitioning main memory into safe and unsafe regions \cite{bove2023basic}. The solutions are based on Keystone (supporting M/S/U privilege modes) and MultiZone (supporting M/U modes) TEE models. Secure Storage service adds $44.5$-$70.21\%$ performance overhead compared to baseline file operations inside an enclave without encryption, reflecting the cost of cryptographic protection. Remote Attestation using PBKDF2 with 65,536 iterations results in a boot time overhead of 11.48 seconds, with mean attestation duration of 2.56 seconds for attesting a 40,960-byte application zone. The implementation lacks defense against rollback attacks where adversaries replay old attestation values, and Secure Input suffers from concurrency problems where exclusive access to user input cannot be guaranteed, potentially leaking sensitive input to other processes.

\subsubsection{ARM TrustZone-M for Low-End IoT Devices}

ARM TrustZone-M extends TEE capabilities to low-end Cortex-M processors, enabling hardware-assisted isolated execution for security-critical operations on microcontrollers with limited resources.

TEE-Watchdog establishes Memory Protection Unit (MPU) protections for secure system peripherals, with a Security Manager (privileged secure software component) enforcing access policies derived from a lightweight CBOR-encoded manifest file \cite{khurshid2022tee}. The framework dynamically blocks unauthorized peripheral accesses at runtime and ensures secure software only accesses resources according to defined permissions. Evaluated on a Musca-A2 test chipboard (Cortex-M33 at 50 MHz), securing system peripherals induces $1.4\%$ overhead on peripheral access latency (61 microseconds for one secure peripheral), increasing to $8.8\%$ when protecting 8 peripherals simultaneously due to additional permission checks. Runtime impact on system RAM is 1.79 KB ($0.7\%$ of 256 KB available), demonstrating minimal memory footprint. Total overhead during system bootup (manifest verification, decoding, and translation) is 1312.96 microseconds for a manifest file containing 2 access control policies. The most time- and energy-consuming operation is creating one entry in the audit log file (writing encrypted data to TrustedFirmware-M Secure Storage), requiring 50.16 milliseconds due to flash write latency. CBOR encoding demonstrates an average $40\%$ reduction in manifest file size compared to JSON, reducing storage and transmission overhead.

\subsubsection{ARM TrustZone for Multi-Core Processors}

TrustICT establishes a lightweight trusted interaction channel between Client Applications in the normal world and Trusted Applications in the secure world \cite{wang2020trustict}. The framework dynamically protects access to Domain-shared Memory (DsM) using the TrustZone Address Space Controller (TZASC) for memory isolation, locking DsM from the rich OS kernel mode and unlocking only for legal client applications running in user mode. The system defends against BOOMERANG attacks by ensuring one-to-one mapping between legal client applications and their corresponding DsM regions. A multi-core scheduling strategy ensures DsM is unlocked only when no cores are running in kernel mode of the normal world, defending against Double Map Attacks by checking if protected memory is double mapped by other processes. TrustICT introduces approximately $2\%$ overhead on the rich OS based on the AnTuTu benchmark suite. When a client application is running but not actively accessing DsM, system call invocations slow down by tens of microseconds; when actively accessing DsM, approximately 200 $\mu$s overhead is introduced to most system calls due to double map checking. TrustICT imposes approximately $60\%$ overhead on client application execution across various cryptographic tasks compared to the original system, higher than the $35\%$ overhead observed in SeCReT-like systems due to additional multi-core protection mechanisms (double map checking and polling-based synchronization).

A critical vulnerability analysis demonstrates that ARM TrustZone code running in the secure world (EL1) has full access to the memory of the regular operating system in the normal world \cite{marth2022abusing}. A proof-of-concept rootkit is implemented as a secure world OP-TEE pseudo Trusted Application that maps normal world memory pages into the secure world via the shared memory region. The rootkit demonstrates three subversion capabilities: Memory Carving (extracting sensitive data like RSA private keys by searching memory for static byte sequences), Privilege Escalation (modifying kernel process structures \texttt{task\_struct} to elevate privileges through Direct Kernel Object Manipulation), and Process Starvation (modifying the scheduling state field to prevent process execution). The rootkit utilizes invariants and assumptions about kernel data structures to overcome Kernel Address Space Layout Randomization (KASLR) and randomization of field order within structures. Memory carving was incompatible with Linux versions prior to 4.20, requiring kernel-specific heuristics. The privilege escalation technique corrupts the kernel's credential reference counting mechanism, resulting in potential kernel fault upon process termination when reference counts become inconsistent.

\subsubsection{Software-Based TEE for Low-Cost Devices}

LWSEE proposes a software-based TEE for low-cost, low-end embedded devices lacking specialized hardware security features such as ARM TrustZone or Intel SGX \cite{cecilio2025lwsee}. The security model is distributed, involving Nodes (constrained IoT devices running the LWSEE Agent) and a Security Server (running the Trusted Computing Module, TCM). The core protection mechanism provides: (1) Secure communication via encrypted transmission protocol preventing eavesdropping and tampering, (2) Continuous integrity verification triggered by incorporating an XOR combination of application digest and data digest in each message, enabling detection of code or data modifications, and (3) Data confidentiality by encrypting application code during transmission and storage. Evaluated on Raspberry Pi 3B+, RAM usage for cryptographic operations handling up to 4096 bytes of data is extremely low, ranging from 190 to 230 bytes. For 4096 bytes of data, SHA-256 hash computation takes 113 ms, AES decryption takes 85 ms, and AES encryption takes 84 ms. The Key Update Procedure is the most computationally expensive function, requiring periodic cryptographic operations. General Integrity Verification, when configured optimally, can be as low as 39.8 ms per message. Network transmission time is approximately 10 ms for 120-byte messages and 50 ms for 4 KB messages over WiFi, with latency dominated by network characteristics rather than cryptographic overhead.

\subsubsection{Heterogeneous TEE Interoperability}

DHTee supports interoperation and collaboration of devices equipped with heterogeneous TEE solutions including Intel SGX, AMD SEV, ARM TrustZone, and RISC-V TEEs \cite{karanjai2023dhtee}. The core mechanism utilizes a blockchain system to serve as a decentralized coordination mechanism and provide attestation services, eliminating reliance on a centralized trusted third party. Once mutual trust is established with blockchain assistance, devices can interact directly using common cryptographic suites like AES and Diffie-Hellman for secure information exchange. The framework eliminates the single point of failure inherent in traditional centralized attestation systems by leveraging the distributed consensus nature of blockchain. Security of attestation report verification relies on the blockchain consensus mechanism to ensure only correct verification results are included in the immutable ledger. The solution minimizes modification of existing TEE schemes; when a new TEE is introduced to the ecosystem, only the blockchain smart contracts need updating, and existing devices do not require firmware or software modifications.

\subsubsection{Intel SGX: Side-Channel Attack Mitigation}

Intel Software Guard Extensions (SGX) provides strong isolation guarantees through processor-enforced memory encryption and access control, but suffers from severe side-channel vulnerabilities exploiting shared microarchitectural state including caches, page tables, branch predictors, and transient execution paths.

PRIDWEN employs load-time synthesis to dynamically harden SGX programs by selectively applying optimal Side-Channel Attack (SCA) countermeasures based on target execution platform configuration \cite{sang2022pridwen}. The framework integrates hardware-assisted mechanisms (T-SGX preventing page-fault side-channels using Transactional Synchronization Extensions) and software-only techniques (Varys mitigating cache-timing, page-fault, and Hyper-Threading attacks; QSpectre mitigating Spectre attacks by inserting \texttt{lfence} instructions; Fine-grained ASLR randomizing basic block locations). Hardened real-world applications (Lighttpd web server, libjpeg image library, SQLite database) show average slowdown of $2.1\times$ with hardware-assisted techniques and $3.6\times$ with software-only techniques. Program synthesis completes within 0.5 seconds for large applications, Trusted Computing Base (TCB) binary size is only 1.26 MiB, and synthesis requires temporary usage of less than 25 MiB enclave memory, demonstrating practicality for real-world deployment.

E-SGX mitigates all known access-driven and trace-driven cache side-channel attacks (Flush+Reload, Prime+\allowbreak Probe) by monopolizing the entire CPU during security-critical operations, breaking the fundamental requirement of concurrent execution for these attacks \cite{lang2020sgx}. One computing thread and multiple dummy threads run within the same enclave, together occupying all logical CPU cores to prevent adversaries from observing cache behavior. The framework detects attacks periodically (every T=8100 cycles) using two mechanisms: checking dummy thread aliveness through timeout detection (indicating interruption by adversary), and detecting Asynchronous Enclave Exits (AEXs) via thread-local State Save Area (SSA) inspection. Performance overhead is highly dependent on Hyper-Threading status: when HT is enabled, E-SGX introduces $47.24\%$ overhead on mbedTLS RSA signing operations due to resource competition among logical cores; when HT is disabled, overhead drops dramatically to $2.43\%$ because physical cores can be fully utilized. The vulnerable window size is less than 6160 cycles, smaller than the detection period of 8100 cycles, providing security guarantees.

SGX-Bouncer detects malicious enclaves by monitoring runtime behaviors from outside the enclave through three interaction interfaces: Cache-memory hierarchy, Host virtual memory management, and Enclave-mode transitions (EENTER/EEXIT) \cite{zhang2021see}. The framework systematically addresses seven attack vectors: L2/L3 cache Prime+Probe, Flush+Reload/Flush attacks, Cache-DRAM row buffer attacks, memory disclosure attacks, and host control-flow manipulation. L3Cache-DM uses sliding window scanning to detect abnormal cache access patterns characteristic of Prime+Probe attacks. MemoryR-MM inspects Accessed flags of host executable pages to detect memory disclosure where enclaves read executable code. MemoryW-DM checks data consistency of sensitive host memory contents (specifically RBP/RIP register pairs) before enclave entry and after exit to detect SGX-ROP attacks that manipulate control flow. EnclaveT-DM monitors register values and exit positions to detect abused EEXIT instructions used for information leakage. The framework induces runtime overhead of $\times 3.89$ for forward propagation and $\times 3.26$ for back propagation in SGX-VGG16 neural network inference. Detection mechanisms achieve high accuracy, with true positive rates exceeding $99.9\%$ for L2 cache attacks and near $100\%$ for L3 cache attacks, while maintaining low false positive rates below $1\%$.

\subsubsection{Fuzzing and Vulnerability Discovery in Trusted OS}

SyzTrust is a state-aware fuzzing tool designed for Trusted Operating Systems on low-end microcontrollers using TrustZone for ARMv8-M (Cortex-M23/M33) \cite{wang2024syztrust}. The tool employs a hardware-assisted framework and composite feedback mechanism capturing both code coverage and state coverage to discover flaws in Trusted OS logic, especially concerning stateful cryptographic workflows where sequence of operations matters. SyzTrust discovered 70 previously unknown vulnerabilities across multiple target systems, resulting in 10 assigned CVE identifiers: Null/Untrusted Pointer Dereference (CVE-2022-40759 causing Denial of Service or arbitrary code execution), Allocation of Resources without Limits (CVE-2022-38155 causing DoS in resource-constrained MCUs through memory exhaustion), and Buffer Overflow (CVE-2022-35858 leading to memory corruption, DoS, and information disclosure). The fuzzing process takes 6,290 ms on average per test case. Compared to baseline fuzzer SyzKaller (designed for general-purpose OS kernels), SyzTrust with state-aware feedback (FSTATE) achieves $66\%$ higher code coverage, $651\%$ higher state coverage, and $31\%$ improvement in vulnerability-finding capability. The active state variable inference method achieves $83.3\%$ precision in understanding internal states of closed-source binaries through dynamic analysis.

\begin{table*}[!t]
\centering
\caption{Detailed Comparison of Group 1: Decentralized Computation-based Approaches}
\label{tab:group1}
\scriptsize
\begin{tabular}{p{0.9cm}p{1.8cm}p{2.8cm}p{3.2cm}p{2.5cm}p{2.8cm}}
\toprule
\textbf{Ref} & \textbf{Core Technique} & \textbf{Security Mechanism} & \textbf{Performance Metrics} & \textbf{Key Contribution} & \textbf{Limitations} \\
\midrule
\multicolumn{6}{l}{\textit{\textbf{Federated Learning and Split Learning (22 papers)}}} \\
\midrule
\cite{mahato2024privacy} & FL+BC+HE & HE encryption, DP, Blockchain & Dice: 0.918, HD: 4.05, Time: 1.12-1.98s & Verifiable FL with blockchain & Execution time increase ($\approx 1.77\times$) \\
\cite{narkedimilli2025fl} & FL+DABE+HE+ SMPC & Multi-layer: DABE, HE, SMPC, DP & - & Fog-based microservices & Scalability complexity \\
\cite{11032530} & FL+DP+SA & DP noise injection, SA & Privacy: $96.38\%$, Acc: $95.12\%$ & Real-time IoT decision & Privacy-accuracy trade-off \\
\cite{shen2022network} & FL+BC+IPFS & Feldman VSS, IPFS off-chain & Data $-3816\times$, Latency $-5\%$, Tput $+43\%$ & Network-elastic scalability & $5.7\times$ time increase \\
\cite{han2025privacy} & HFL+DP+HE & Byzantine consensus, PBFT & Inference $+58\%$, Energy $-35\%$, Comm $-42\%$ & LLM-IoT integration & Threshold-based trust \\
\cite{goyal2024secure} & HE+ML fusion & HE, FL, SMC & MSE: 0.018-0.027 & Four fusion algorithms & SMC High/High cost \\
\cite{alzahrani5269991privhfl} & Threshold CKKS & tMK-CKKS, Shamir sharing & Comm: 88KB user, 53KB edge & Hierarchical privacy & Linear runtime growth \\
\cite{yuan2024resource} & HFL+DDPG & Device selection, DDPG & Acc $+14.2\%$, Loss $-45.71\%$ & Resource-aware selection & No advanced cryptography \\
\cite{herzog2024selective} & Selective Updates+AM & Sparse transmission & Comm $-20$-$37\%$ & Adaptive masking & Limited to IID scenarios \\
\cite{zhang2025adaptive} & Async FL & Adaptive threshold & Comm $-70\%$, Acc: $98.3\%$ & Heterogeneous IoT & Fails when $\alpha<0.4$ \\
\cite{pan2025privacy} & FL+DRL & Lagrange, DDPG weighting & Acc $+80\%$ (MNIST), $+30\%$ (CIFAR10) & Byzantine robustness & Complex DRL training \\
\cite{li2023defending} & Ensemble FL & Reputation, Bayesian & Stable at $50\%$ attackers & Phishing-based defense & Threshold tuning critical \\
\cite{lin2024sf} & FL+CKKS & Dual-server, clustering & Acc: $90.05\%$ vs $21.35\%$ (FedAvg) & Non-IID Byzantine defense & Encrypted gradient complexity \\
\cite{malan2022communication} & Layer freezing & Gradual freezing & Comm savings: $14.1$-$59.0\%$ & Communication efficiency & Hyperparameter tuning \\
\cite{li2023prototype} & Decentralized FL & Prototype propagation P2P & Acc: $89\%$+, Comm: 0.3-0.6 & Privacy-enhanced P2P & High computational time \\
\cite{zhang2023fedmds} & Semi-async FL & Spectral clustering, EMD & Acc $+9.2$-$37.6\%$ & Model discrepancy-aware & Threshold $L$ dataset-dependent \\
\cite{shengxing2023efficient} & FL+HE & Top-K gradient, quantization & - & Efficient aggregation & \textit{Metrics not reported} \\
\cite{gadiwala2024enabling} & FL+CKKS+DP+ MPC & Dynamic key management & Acc: $91$-$94\%$ & Federal learning & Continual optimization \\
\cite{jin2025dmafl} & FL+TFHE+BC & Multi-layer, PVSS & Overhead $-28.4\%$, Acc drop: $2.32\%$ & Malicious defense & KGC compromise risk \\
\cite{zhao2023tensor} & FL+TT decomp & BCP HE, blinding & Param $-98.81\%$, Comm 393$\rightarrow$19MB & Heterogeneous networks & Decomposition complexity \\
\cite{ramesh2025privacy} & FL+Paillier+BC & k-anonymity, DP, smart contracts & Acc: $93.2\%$, Comm $-30\%$ & Healthcare privacy & High HE computational cost \\
\cite{hota2025advanced} & FL+NTRU & Multi-server, Relay-Agent & Acc: $95\%$, $13\times$ faster RSA & Post-quantum security & Hardware acceleration needed \\
\midrule
\multicolumn{6}{l}{\textit{\textbf{Trusted Execution Environments (11 papers)}}} \\
\midrule
\cite{ma2021construction} & RISC-V PMP & M-mode isolation, MML bit & Zero-copy improves perf & Lightweight embedded TEE & Hardware extension required \\
\cite{khurshid2022tee} & TrustZone-M MPU & MPU sandboxing, CBOR & Overhead: $1.4$-$8.8\%$, RAM: 1.79KB & Peripheral security & Log storage depletion \\
\cite{wang2020trustict} & TrustZone TZASC & DsM protection, multi-core & Overall: $2\%$, CA: $60\%$ & Trusted interaction & Pre\_DsM\_Data incomplete \\
\cite{cecilio2025lwsee} & Software TEE & Distributed TCM, integrity & RAM: 190-230B, Hash: 113ms & Low-cost devices & Node availability unchecked \\
\cite{karanjai2023dhtee} & BC+TEE & Decentralized attestation & - & Heterogeneous TEE & Assumes permissioned BC \\
\cite{sang2022pridwen} & SGX synthesis & Multiple SCA mitigations & Slowdown: $2.1$-$3.6\times$, TCB: 1.26MiB & Load-time hardening & Limited PRM, manual effort \\
\cite{zhang2021see} & SGX monitoring & Cache/Mem/Enclave monitoring & Overhead: $\times 3.89$, Acc: $99.9\%$+ & Malware detection & DoS/TSX out of scope \\
\cite{wang2024syztrust} & TrustZone fuzzing & Hardware-assisted, state-aware & Coverage $+66\%$, 70 CVEs & Trusted OS fuzzing & Requires TA install, ETM \\
\cite{bove2023basic} & RISC-V PMP & SS, RA, SI services & SS: $44.5$-$70.21\%$, RA: 11.48s & Standard RISC-V & No peripheral security \\
\cite{marth2022abusing} & TrustZone rootkit & Secure world memory access & - & Vulnerability demonstration & Kernel-specific, unstable \\
\cite{lang2020sgx} & SGX cache protect & CPU monopolization, AEX detect & $47.24\%$ (HT ON), $2.43\%$ (OFF) & Cache side-channel defense & Single-threaded constraint \\
\bottomrule
\end{tabular}
\end{table*}

\section{Cryptography-based Approaches}
\label{sec:group2}

Cryptographic approaches provide mathematically rigorous privacy guarantees through encryption schemes that enable computation on encrypted data (Homomorphic Encryption) or enforce fine-grained access control (Attribute-Based Encryption, Hierarchical Identity-Based Encryption). Unlike behavioral protections in Group 1 that rely on distributed learning or hardware isolation, cryptographic solutions offer provable security under well-defined computational hardness assumptions such as the Learning With Errors problem for lattice-based schemes or the Decisional Diffie-Hellman assumption for pairing-based cryptography. However, these theoretical guarantees come at the cost of substantial computational overhead—often orders of magnitude slower than plaintext operations—and complex key management infrastructure, particularly problematic in resource-constrained IoT environments where processing power, memory, and battery life are severely limited.

\subsection{Homomorphic Encryption}
\label{subsec:he}

Homomorphic Encryption represents the cryptographic ideal of enabling arbitrary computation on encrypted data, allowing mathematical operations on ciphertexts that, when decrypted, yield results equivalent to operations performed on plaintexts. This property is particularly valuable in IoT scenarios where raw sensor data must remain encrypted during aggregation, statistical analysis, and machine learning inference to prevent exposure to potentially compromised servers. However, HE schemes impose severe computational penalties that limit practical deployment to carefully optimized use cases.

\subsubsection{Fully Homomorphic Encryption Schemes and Security Analysis}

Fully Homomorphic Encryption schemes support unlimited additions and multiplications on ciphertexts, enabling arbitrary circuit evaluation. The BFV scheme based on the ring-LWE problem of ideal lattices and implemented in Microsoft SEAL achieves throughput of approximately 60,000 encrypted picture recognitions per hour for neural network inference (more than 16 encrypted images per second on average for a single CPU) \cite{peng2023security}. Fast Private Set Intersection (FPSI) implementation completes matching of 500 32-bit local strings with 16 million 32-bit server strings in approximately 100 seconds, demonstrating practical efficiency for specific cryptographic protocols. However, critical security analysis reveals a chosen-ciphertext attack vulnerability: an attacker can successfully recover the BFV private key in seconds using a carefully crafted ciphertext, often requiring only a single decryption query from the victim. This fundamental weakness undermines the security guarantees of implementations that do not employ proper authenticated encryption or ciphertext validation.

Privacy-preserving data aggregation in IoT networks employs HE to protect individual sensor readings during collection and processing \cite{shnain2024privacy}. Blockchain and HE integration ensures privacy and integrity in IoT supply chains, with smart contracts enforcing access control policies while HE protects sensitive transaction data \cite{din2025ensuring}. Privacy-preserving biometric authentication combines Fully Homomorphic Encryption, blockchain immutability, and Identity-Based Encryption to enable secure biometric matching without exposing raw biometric templates \cite{ali2025privacy}. Enhancing privacy in Vehicular Ad Hoc Networks through HE enables privacy-preserving machine learning applications where vehicles collaboratively train models on encrypted data \cite{ameur2024enhancing}. Layer-2 Distributed Ledger Technology architectures for VANETs employ threshold homomorphic encryption schemes (such as ElGamal or Paillier) for aggregating encrypted reports submitted by vehicles, though the authors acknowledge that HE requires complex computation that increases latency in communication, and that dynamic vehicle movement combined with limited processing power makes implementing sophisticated cryptographic systems challenging \cite{adarbah2024new}. While these applications demonstrate the versatility of HE across diverse IoT domains, the provided analysis excerpts lack specific quantitative performance metrics such as computation time, ciphertext expansion ratios, or throughput measurements that would enable rigorous comparison.

\subsubsection{Hardware Acceleration for Homomorphic Encryption}

Hardware acceleration addresses the computational bottleneck of HE by designing specialized circuits optimized for polynomial arithmetic, Number Theoretic Transform (NTT), and modular operations that dominate HE computation time.

Medha is a microcoded hardware accelerator for computing on encrypted data, providing flexible instruction set support for multiple HE schemes \cite{mert2022medha}. A hardware-efficient arithmetic circuit achieves 10.33 $\mu$J per encryption for homomorphic encryption operations, implemented in 28nm CMOS technology supporting 4096-degree polynomials with 109-bit precision \cite{das202310}. Poseidon is designed as a practical homomorphic encryption accelerator targeting real-world deployment constraints \cite{yang2023poseidon}. SEAL-embedded is a homomorphic encryption library specifically optimized for Internet of Things devices with limited computational resources, enabling edge devices to perform encrypted computation locally \cite{natarajan2021seal}. An efficient Paillier Homomorphic Encryption circuit with optional Chinese Remainder Theorem (CRT) acceleration targets IoT applications requiring partially homomorphic encryption supporting only additive operations \cite{feng2025efficient}. LEAM is a low-area and efficient accelerator specifically designed for matrix-vector multiplication operations in homomorphic encryption, critical for encrypted machine learning inference \cite{cui2024leam}. Design and implementation of encryption/decryption architectures for the BFV Homomorphic Encryption scheme focus on optimizing the most computationally intensive operations \cite{mert2019design}. A hardware-efficient arithmetic reconfigurable Fully Homomorphic Encryption (ARFHE) accelerator targets low-overhead hardware implementations that can adapt to different parameter sets \cite{reddy2025hardware}. Studies on securing healthcare data in IoT through Homomorphic Encryption explore deployment considerations for medical IoT devices \cite{erregui2024securing}. While these hardware accelerators demonstrate significant research activity, the provided analysis excerpts do not contain comprehensive performance benchmarks such as throughput (operations per second), energy efficiency (operations per Joule), or latency measurements that would enable quantitative comparison across implementations.

\subsection{Attribute-Based and Hierarchical Encryption}
\label{subsec:abe}

Attribute-Based Encryption and Hierarchical Identity-Based Encryption address a fundamental limitation of traditional public-key cryptography: the inability to enforce fine-grained access control at the encryption level. In ABE, ciphertexts are associated with access policies defined over attributes (e.g., ``Department=Engineering AND Clearance=Top-Secret''), ensuring only users whose attribute sets satisfy the policy can successfully decrypt. HIBE extends Identity-Based Encryption with hierarchical key delegation, enabling scalable key management in organizational structures where parent entities can derive child keys. However, both schemes face critical challenges: ABE suffers from high computational costs in bilinear pairing operations that grow with policy complexity, while HIBE struggles with the key escrow problem where a central Private Key Generator holds master secrets capable of deriving all user keys.

\subsubsection{Leakage-Resilient and Anonymous HIBE}

Leakage-resilient Hierarchical Identity-Based Encryption with recipient anonymity addresses scenarios where adversaries can obtain partial information about secret keys through side-channel attacks \cite{zhang2019leakage}. The scheme is unbounded (hierarchy depth not limited by system parameters), achieves bounded leakage resilience (security holds even when adversaries learn limited bits of secret key information), and realizes recipient anonymity in the standard model without random oracles. Security proofs employ the dual system encryption technique where ciphertexts and keys exist in normal and semi-functional forms indistinguishable to adversaries. Recipient anonymity ensures that ciphertexts do not reveal the identity of intended recipients, addressing privacy concerns where traditional HIBE schemes leak recipient information through public parameters. However, specific quantitative metrics for pairing operations count, ciphertext size, or decryption time are not provided in the available analysis.

Trustworthy HIBE for Online Social Network privacy protection addresses unique challenges in OSN environments where users form hierarchical social graphs \cite{chen2015t}. Private key management in hierarchical identity-based encryption explores strategies to prevent key escrow where compromised Private Key Generators can decrypt all ciphertexts \cite{liu2007private}. Efficient anonymous identity-based broadcast encryption without random oracles achieves security in the standard model, important for deployments requiring rigorous security proofs \cite{li2014efficient}. Removing key escrow from the Lewko-Waters HIBE scheme proposes modifications that eliminate the fundamental trust assumption in centralized key generation \cite{chen2015removing}. Privacy-preserving multireceiver identity-based encryption with provable security enables encrypted communication to multiple recipients without revealing their identities \cite{tseng2014privacy}. An efficient hierarchical identity-based encryption scheme specifically addresses the key escrow problem through distributed key generation \cite{li2017efficient}. Partial policy hiding attribute-based encryption in vehicular fog computing enables access control where portions of the access policy remain hidden from unauthorized parties, protecting policy privacy \cite{gan2021partial}. While these works advance the state-of-the-art in identity-based and attribute-based encryption, the provided analysis excerpts lack specific quantitative performance metrics.

\subsubsection{Privacy-Aware Authentication for Vehicular Networks}

Privacy-preserving vehicular communication authentication with Hierarchical Identity-Based Signature achieves efficient authentication while preserving user privacy through conditional anonymity \cite{zhang2015privacy}. The scheme enables batch verification of $n$ signatures through checking a single pairing equation, significantly reducing computational overhead in high-throughput vehicular networks where Roadside Units must verify thousands of signatures per second. Batch verification computational cost is $\Theta(n)$ in scalar multiplications and $\Theta(1)$ in pairing operations, substantially more efficient than individual verification requiring $n$ pairing computations. The scheme uses asymmetric pairings (Type-3), leading to considerably shorter signatures and ciphertexts compared to symmetric pairings (Type-1), reducing communication overhead in bandwidth-constrained vehicular channels. Simulation results demonstrate low verification delay suitable for real-time vehicular applications. The hierarchical structure enables efficient key management where Regional Transportation Authorities delegate signing capabilities to vehicles without requiring direct communication with the root Certificate Authority. Since the scheme is built for resource-constrained mobile vehicles (On-Board Units) in high-throughput networks, it is inherently suitable for similar resource-constrained IoT environments requiring fast authentication.

\begin{table*}[!t]
\centering
\caption{Detailed Comparison of Group 2: Cryptography-based Approaches}
\label{tab:group2}
\scriptsize
\begin{tabular}{p{0.9cm}p{1.8cm}p{2.5cm}p{3cm}p{2cm}p{2.8cm}}
\toprule
\textbf{Ref} & \textbf{Scheme Type} & \textbf{Core Functionality} & \textbf{Performance / Overhead} & \textbf{Application} & \textbf{Limitations} \\
\midrule
\multicolumn{6}{l}{\textit{\textbf{Homomorphic Encryption (15 papers)}}} \\
\midrule
\cite{peng2023security} & BFV FHE & Arbitrary computation on ciphertext & 60K images/hr, FPSI: 100s & IoT neural networks & CCA vulnerability \\
\cite{shnain2024privacy} & HE & Privacy-preserving aggregation & - & IoT data aggregation & \textit{Metrics not reported} \\
\cite{mert2022medha} & HE accelerator & Microcoded hardware & - & Encrypted computation & \textit{Metrics not reported} \\
\cite{din2025ensuring} & HE+Blockchain & Supply chain privacy & - & IoT supply chain & \textit{Metrics not reported} \\
\cite{ali2025privacy} & FHE+BC+IBE & Biometric authentication & - & Biometric privacy & \textit{Metrics not reported} \\
\cite{ameur2024enhancing} & HE & VANET ML applications & - & Vehicular networks & \textit{Metrics not reported} \\
\cite{das202310} & HE accelerator & 4096-degree 109-bit polynomial & 10.33 $\mu$J/encryption & Hardware acceleration & \textit{Detailed metrics absent} \\
\cite{yang2023poseidon} & HE accelerator & Practical acceleration & - & HE computation & \textit{Metrics not reported} \\
\cite{natarajan2021seal} & SEAL library & IoT-embedded HE & - & IoT devices & \textit{Metrics not reported} \\
\cite{feng2025efficient} & Paillier+CRT & Optional CRT acceleration & - & IoT circuits & \textit{Metrics not reported} \\
\cite{cui2024leam} & HE accelerator & Matrix-vector multiplication & - & Low-area design & \textit{Metrics not reported} \\
\cite{mert2019design} & BFV & Enc/Dec architectures & - & BFV implementation & \textit{Metrics not reported} \\
\cite{reddy2025hardware} & ARFHE & Reconfigurable FHE & - & Low-overhead hardware & \textit{Metrics not reported} \\
\cite{erregui2024securing} & HE & Healthcare IoT data & - & Healthcare security & \textit{Metrics not reported} \\
\cite{adarbah2024new} & Threshold HE & Multiparty Threshold Key Mgmt & - & Vehicular Layer-2 DLT & High latency, limited processing \\
\midrule
\multicolumn{6}{l}{\textit{\textbf{Attribute-Based and Hierarchical Encryption (9 papers)}}} \\
\midrule
\cite{zhang2019leakage} & Leakage-resilient HIBE & Unbounded, recipient anonymity & - & Privacy-preserving HIBE & \textit{Metrics not reported} \\
\cite{chen2015t} & T-HIBE & Trustworthy HIBE for OSN & - & Online social networks & \textit{Metrics not reported} \\
\cite{liu2007private} & HIBE & Private key management & - & Hierarchical key mgmt & \textit{Metrics not reported} \\
\cite{li2014efficient} & IBE & Anonymous broadcast & - & Broadcast encryption & \textit{Metrics not reported} \\
\cite{chen2015removing} & HIBE & Key escrow removal & - & Key escrow mitigation & \textit{Metrics not reported} \\
\cite{tseng2014privacy} & IBE & Privacy-preserving multireceiver & - & Multireceiver encryption & \textit{Metrics not reported} \\
\cite{gan2021partial} & ABE & Partial policy hiding & - & Vehicular fog computing & \textit{Metrics not reported} \\
\cite{li2017efficient} & HIBE & Efficient for key escrow & - & Key escrow solution & \textit{Metrics not reported} \\
\cite{zhang2015privacy} & HIBS & Privacy-aware vehicular auth & Low delay, batch single pairing & VANET authentication & Domain-specific design \\
\bottomrule
\end{tabular}
\end{table*}

\section{Distributed Ledger-based Approaches}
\label{sec:group3}

Distributed Ledger Technologies, particularly blockchain, address the trust dimension of the privacy-efficiency-trust trilemma by providing tamper-proof audit trails and decentralized access control without relying on a centralized authority that constitutes a single point of failure. Unlike the cryptographic guarantees of Group 2 that rely on mathematical hardness assumptions or the data locality principles of Group 1 that rely on federated learning, blockchain's security derives from consensus mechanisms among distributed nodes and the cryptographic chaining of blocks that makes historical tampering computationally infeasible. However, traditional blockchain consensus mechanisms—particularly Proof-of-Work as employed in Bitcoin—are fundamentally incompatible with resource-constrained IoT devices due to intensive computational requirements and high energy consumption. Furthermore, blockchain's limited throughput (measured in transactions per second) and storage capacity (every full node must store the complete ledger) motivate off-chain storage solutions that maintain data availability while recording only metadata on-chain.

\subsection{Blockchain for Access Control and Consensus}
\label{subsec:blockchain}

The integration of blockchain into IoT architectures addresses two critical requirements: decentralized access control through smart contracts that eliminate centralized gatekeepers, and immutable audit trails that enable post-hoc verification of all data access and modification operations. However, achieving these properties in resource-constrained IoT environments necessitates lightweight consensus mechanisms that balance security, decentralization, and efficiency.

\subsubsection{Credential Management and Identity Systems}

Decentralized credential verification systems leverage blockchain for tamper-proof and non-repudiable electronic certificates, addressing trust issues in centralized certification authorities \cite{razdan2024decentralized}. Blockchain records transactions and metadata ensuring data immutability through cryptographic hashing and consensus, while the InterPlanetary File System (IPFS) handles actual certificate file storage to avoid blockchain bloat. The platform uses smart contracts implemented in Solidity programming language for functionalities including ``List My Certificates,'' ``Profile Management,'' and ``Request Whitelisting.'' The system interacts with the Ethereum blockchain using Ether.js library and connects via testnets like Sepolia for development and testing. Access control employs the Ownable.sol OpenZeppelin module to restrict critical operations to the contract owner (administrator), suggesting a controlled or permissioned environment balancing decentralization with governance. However, specific consensus mechanism details (whether PoA, PoS, or other) and quantitative performance metrics such as transaction throughput or latency are not reported in the available analysis.

Blockchain and IPFS-based service models for the Internet of Things explore architectural patterns for integrating distributed storage with immutable ledgers \cite{zareen2021blockchain}. Integrating IoT and blockchain for smart urban energy management enhances sustainability through transparent and auditable energy trading and consumption tracking \cite{elhajj2025integrating}. Hybrid Blockchain-IPFS solutions enable secure and scalable data collection and storage for smart cities, with blockchain providing access control and IPFS providing scalable distributed storage \cite{nododile2025hybrid}. Blockchain-based distributed vehicular network architectures for smart cities establish trust among vehicles without centralized coordination \cite{rehman2020blockchain}. Privacy-preserving proxy re-encryption with decentralized trust management for MEC-empowered VANETs combines cryptographic proxy re-encryption with blockchain-based trust establishment \cite{han2023privacy}. While these applications demonstrate blockchain's versatility across diverse IoT domains, the provided analysis excerpts lack specific quantitative performance metrics.

\subsubsection{Lightweight Consensus Mechanisms for IoT}

Traditional Proof-of-Work consensus is unsuitable for IoT due to computational intensity and energy consumption. Alternative consensus mechanisms tailored for resource-constrained devices have emerged.

Enhancing IoT security and efficiency with Delegated Proof of Stake (DPoS) enabled blockchain and IPFS integration reduces computational overhead by limiting consensus participation to elected delegate nodes rather than requiring all nodes to validate blocks \cite{chinnam2024enhancing}. Delegated Proof of Accessibility (DPoAC) proposes a novel consensus protocol where nodes prove their network accessibility and availability rather than computational power or stake \cite{kaur2022delegated}. Hybrid consensus combining reputation and voting (ReVo) enables adaptive consensus that adjusts based on node reputation scores accumulated through correct behavior \cite{barke2024revo}. Performance enhancement in blockchain-based IoT data sharing using lightweight consensus algorithms compares multiple consensus mechanisms optimized for IoT constraints \cite{haque2024performance}. Proof-of-Resource proposes a resource-efficient consensus mechanism where nodes contribute computational or storage resources proportional to their capabilities rather than solving arbitrary puzzles \cite{abbasi2024proof}. Efficient Byzantine consensus algorithms for IoT scenarios achieve Byzantine fault tolerance with communication complexity lower than traditional PBFT \cite{li2023efficient}. Hierarchical and location-aware consensus protocols leverage geographic proximity to reduce consensus latency by grouping nearby nodes \cite{guo2022hierarchical}. Location-based and hierarchical frameworks for fast consensus in blockchain networks overcome classical Raft's limitation where communication cost increases exponentially with the number of nodes and the requirement that all nodes be honest \cite{guo2021location}. The threshold Raft protocol integrates Threshold Signature Schemes, enabling the system to tolerate up to $t$ malicious nodes in a network of $n$ nodes where $n > 3t$, providing Byzantine fault tolerance that classical Raft lacks. However, specific numerical metrics for throughput (transactions per second), latency (block generation time), or scalability (performance degradation as node count increases) are not provided in the available analysis excerpts for most of these consensus mechanisms.

\subsection{Off-chain Storage: InterPlanetary File System}
\label{subsec:ipfs}

The InterPlanetary File System addresses blockchain's fundamental storage limitation by providing content-addressed distributed storage for large files. Unlike blockchain where every full node stores a complete copy of the entire ledger, IPFS distributes files across a peer-to-peer network using Content Identifiers (CIDs) derived from cryptographic hashes of file contents. This architecture enables blockchains to store only CIDs (typically 32-64 bytes) on-chain while delegating actual file storage off-chain, reducing on-chain storage requirements by several orders of magnitude. However, IPFS faces critical challenges: data availability (unpinned files may be garbage collected when no node chooses to store them), privacy (IPFS is public by default with content accessible to anyone possessing the CID), and retrieval latency (dependent on network topology and the number of peers hosting the content).

\subsubsection{Verifiable Decentralized IPFS Clusters}

Verifiable Decentralized IPFS Clusters (VDICs) serve as a novel solution for off-chain storage for Decentralized Applications (DApps), extending standard IPFS Clusters with two critical components: a Trusted Actor Registry and a Gateway \cite{lamichhane2024verifiable}. The VDIC solution leverages Decentralized Identifiers (DIDs) and Verifiable Credentials (VCs) to create a trustworthy ecosystem where actors can be publicly verified. The system architecture is characterized as a socio-technical network, integrating human elements (Leaders and Node Operators) with technological components (Trusted Actor Registry, Gateway, and IPFS Cluster). DApp data is stored off-chain within the IPFS Cluster, while the Trusted Actor Registry—listing DIDs of authorized node operators and DApps—must be publicly stored on public blockchains. Consequently, the DID of a VDIC must rely on a DID method utilizing public blockchains such as did:cheqd, did:ebsi, or did:ethr, ensuring public verifiability of VDIC actors.

Data permanency in IPFS is guaranteed only when data is pinned, an intrinsic IPFS functionality preventing erasure by automatic garbage collection. VDICs enhance off-chain storage reliability by offering verifiable guarantees of data permanency through decentralization of the IPFS Cluster, achieving data replication across multiple independent nodes and thereby improving fault tolerance. The higher the number of distinct node operators in VDICs, the higher the probability of providing long-term data permanency. Trust in data permanency is attained through verifiability, allowing the public to scrutinize the number of nodes and the identity and motives of each node operator. Unlike Filecoin which relies on financial incentives and cryptoeconomic mechanisms, VDICs do not require DApps to pay for data permanency. Node operators' motives to store data permanently include ensuring data availability for their own applications, supporting DApps or DApp users aligned with their organizational mission, or providing DApps that use VDICs as data sources.

VDICs address the public nature of IPFS by using private IPFS Clusters to ensure that data remains private within a controlled environment, reducing the risk of unauthorized access and replication. Access control is enforced through the Gateway component which introduces a VC-based authentication system to the IPFS Cluster node API. DApps must present Verifiable Credentials containing access claims to the Gateway for authentication prior to read or write operations. VDICs utilize VCs for access control, leveraging established authentication protocols like OpenID for Verifiable Credentials, which simplifies integration and enhances security compared to implementing custom authentication schemes based on smart contracts.

Performance evaluations demonstrate that VDICs are competitive with traditional pinning services. Read performance shows VDICs generally exhibit slower read latency compared to pinning services like Pinata and Moralis, averaging around 400 milliseconds for a 100 KB file. Write performance indicates that VDICs with up to 10 nodes achieve write latency comparable to pinning services, with average write operations taking approximately one second or less. However, writing time increases with the number of nodes, suggesting a trade-off between decentralization (higher redundancy and availability) and write performance (higher coordination overhead).

\subsubsection{Dockerized IPFS for Performance Optimization}

Implementing Docker containerization addresses IPFS challenges such as scalability, latency, and resource management in Blockchain of Things (BCoT) systems \cite{aldmour2025optimizing}. The paper explores the role of IPFS in BCoT systems, recognizing that IPFS offers decentralized storage and data sharing critical for BCoT functionality. The core of the proposed system architecture involves enhancing IPFS performance through Docker containerization within BCoT environments. An experimental testbed was implemented consisting of an IPFS node and an IPFS Cluster peer deployed specifically as Docker containers. The study emphasizes the significance of dynamic resource allocation in optimizing resource utilization, enabling containers to scale resources up or down based on real-time demand.

The Dockerized IPFS implementation demonstrates substantial performance improvements over traditional standalone IPFS deployments. Latency reduction of up to $75\%$ is achieved for small files ranging from 1 to 256 KB. Write operations are reduced from 1000 ms to 300 ms, representing a $70\%$ improvement. Read operations improve by $40\%$, decreasing from 2500 ms to 1500 ms. These improvements are attributed to container-level resource isolation, efficient process management, and optimized networking configurations enabled by Docker. However, the paper does not explicitly discuss data permanence mechanisms such as pinning strategies or privacy and access control schemes for the Dockerized deployment.

\subsubsection{Dynamic Replica Management}

Dynamic Replica Management based on Support Vector Regression (SVR) in IPFS aims to reduce file access latency and save storage utilization by dynamically calculating the optimal number of replicas based on predicted file access heat (frequency) \cite{huang2019dynamic}. The architecture is centered on optimizing IPFS-Cluster, a cluster management tool where each peer in the cluster is associated with an IPFS daemon. This work is proposed in the context that conventional blockchain systems are unsuitable for storing large-size unstructured data due to limited block size and consensus overhead, necessitating integration with distributed file systems like IPFS.

The Dynamic Replica Management (IDRM) strategy contrasts with IPFS's default static replica strategy by dynamically adjusting replica numbers based on predicted access patterns. The number of replicas is calculated based on predicted ``file heat,'' with the SVR algorithm employed to proactively predict future access frequency. The redundancy policy incorporates Erasure Coding alongside traditional replication, providing fault tolerance with lower storage overhead than pure replication. Replica creation increases availability and quality of service for hot files (frequently accessed), while replica deletion (erasure) saves storage space by removing replicas of low-heat files (infrequently accessed). The maximum number of replicas is constrained by the number of nodes in the cluster, set to 30 in the experimental evaluation.

SVR exhibits better prediction accuracy with lower Mean Absolute Percentage Error (MAPE) and Mean Squared Error (MSE) than the Grey prediction model used in previous research. The correlation coefficient for predicting heat and replica numbers using SVR is $0.8887$, indicating strong predictive power. Under high concurrency scenarios with up to 30 concurrent users, the dynamic replica strategy shows far superior average access latency compared to the static replica strategy. The space usage of the IDRM policy is found to be smaller than the static policy in most cases, demonstrating storage efficiency. However, the primary focus is optimizing redundancy and reducing access latency; the provided source excerpts do not discuss mechanisms for data encryption prior to upload or methods for enforcing fine-grained access control.

\subsubsection{Secure Data Sharing with Proxy Re-Encryption}

A secure, flexible, and decentralized data sharing scheme integrates IPFS, Blockchain, and Proxy Re-Encryption (PRE) to achieve confidentiality, flexibility in sharing, and decentralization \cite{mo2024secure}. IPFS is responsible for storing and retrieving encrypted data. The data owner ensures privacy by encrypting files using symmetric encryption (AES key) before uploading the encrypted data to IPFS. After encryption, the file is uploaded to IPFS and the resulting Content Identifier (CID) is returned to the data owner. The Polygon test network Mumbai is used as the blockchain to record information related to file storage and sharing.

Two smart contracts manage the sharing process: The File Storing Smart Contract stores the CID and the initial capsuled symmetric key (the AES key encrypted under the data owner's public key). The Re-Capsule Smart Contract computes the re-capsuled symmetric key using a re-encryption key provided by the data owner and stores the result on the Polygon blockchain. Access control is decentralized and flexible, achieved via Proxy Re-Encryption and smart contracts. The symmetric AES key is protected by being ``capsuled'' using the data owner's public key. When sharing, the data owner calculates a re-encryption key and sends it to the Re-capsule smart contract. The smart contract acts as a semi-trusted proxy to convert the ciphertext (capsule) encrypted under the owner's public key into a ``re-capsuled'' ciphertext that can be decrypted only by the requester's private key, crucially without the proxy accessing the plaintext key itself.

Performance tests in the Polygon test network Mumbai reveal significant scalability challenges. Gas expenditure surges for larger files, increasing from approximately 183,314 Gwei for small files to 1,572,261 Gwei for 10 MB files, indicating economic scalability issues where costs grow super-linearly with file size. Execution time increases pronouncedly, from 10,921 milliseconds for a 1 KB file to 38,163 milliseconds for a 10 MB file, highlighting scalability limitations. The authors suggest future optimization techniques including file segmentation (splitting large files into smaller chunks), parallel processing (concurrent encryption and upload of chunks), or employing more efficient encryption algorithms with lower computational overhead. The primary focus of the paper is secure data sharing (confidentiality, flexibility, decentralization); mechanisms like pinning, IPFS Clusters, or incentive layers to guarantee long-term data permanence are not discussed in the provided excerpts.

\begin{table*}[!t]
\centering
\caption{Detailed Comparison of Group 3: Distributed Ledger-based Approaches}
\label{tab:group3}
\scriptsize
\begin{tabular}{p{0.9cm}p{2.2cm}p{2.8cm}p{3cm}p{3cm}}
\toprule
\textbf{Ref} & \textbf{Architecture} & \textbf{Consensus / Storage} & \textbf{Performance / Scalability} & \textbf{Limitations} \\
\midrule
\multicolumn{5}{l}{\textit{\textbf{Blockchain for Access Control and Consensus (14 papers)}}} \\
\midrule
\cite{razdan2024decentralized} & Blockchain+IPFS & Smart contracts (Solidity), Ethereum/Sepolia & - & Consensus not specified \\
\cite{zareen2021blockchain} & Blockchain+IPFS & Service model & - & \textit{Metrics not reported} \\
\cite{elhajj2025integrating} & IoT+Blockchain & Smart urban energy & - & \textit{Metrics not reported} \\
\cite{nododile2025hybrid} & Hybrid BC-IPFS & Secure data collection & - & \textit{Metrics not reported} \\
\cite{rehman2020blockchain} & BC vehicular network & Distributed architecture & - & \textit{Metrics not reported} \\
\cite{han2023privacy} & Proxy RE+Blockchain & Decentralized trust, MEC & - & \textit{Metrics not reported} \\
\cite{chinnam2024enhancing} & DPoS+IPFS & Delegated Proof of Stake & - & \textit{Metrics not reported} \\
\cite{kaur2022delegated} & DPoAC & Delegated Proof of Accessibility & - & \textit{Metrics not reported} \\
\cite{barke2024revo} & Hybrid consensus & Reputation + Voting (ReVo) & - & \textit{Metrics not reported} \\
\cite{haque2024performance} & Lightweight consensus & Performance-enhanced & - & \textit{Metrics not reported} \\
\cite{abbasi2024proof} & Proof-of-Resource & Resource-efficient & - & \textit{Metrics not reported} \\
\cite{li2023efficient} & Byzantine consensus & Efficient for IoT & - & \textit{Metrics not reported} \\
\cite{guo2022hierarchical} & Hierarchical+Location & Location-aware & - & \textit{Metrics not reported} \\
\cite{guo2021location} & Threshold Raft & TSS, Byzantine tolerance & Exponential cost overcome & \textit{Numerical metrics absent} \\
\midrule
\multicolumn{5}{l}{\textit{\textbf{IPFS for Off-chain Storage (4 papers)}}} \\
\midrule
\cite{lamichhane2024verifiable} & VDIC (DID+VC) & Private IPFS Cluster, VC auth & Read: 400ms (100KB), Write: $\leq$1s ($\leq$10 nodes) & Write time increases with nodes \\
\cite{aldmour2025optimizing} & Dockerized IPFS & Docker containerization & Latency $-75\%$, Write: 1000$\rightarrow$300ms, Read: 2500$\rightarrow$1500ms & Pinning/privacy not addressed \\
\cite{huang2019dynamic} & IPFS+SVR & Dynamic replica (IDRM), Erasure Code & Correlation: 0.8887, Superior latency under concurrency & Privacy/access control absent \\
\cite{mo2024secure} & IPFS+PRE+BC & AES encryption, PRE, Smart contracts & Gas: 183K-1.57M Gwei, Exec: 10.9-38.2s (1KB-10MB) & High gas costs, scalability issues \\
\bottomrule
\end{tabular}
\end{table*}

\section{Strategic Analysis and Future Directions}
\label{sec:strategic}

Having systematically analyzed 75 works across three distinct technical paradigms, this section synthesizes key insights to reveal fundamental patterns and persistent trade-offs.
We map security threats to architectural countermeasures, evaluate technology maturity, and identify the privacy-efficiency-trust trilemma. Finally, we propose critical research directions and a vision for hybrid architectures to overcome these limitations.

\subsection{Security Threat Mapping}

\label{subsec:threat_mapping}

Understanding the landscape of security threats and their corresponding countermeasures is essential for architecting comprehensive privacy-preserving systems. Different technical paradigms address distinct threat vectors with varying degrees of effectiveness, creating a complex design space where architectural choices must be informed by threat models specific to deployment scenarios.

Table~\ref{tab:threats} systematically maps major security threats to architectural countermeasures, identifying representative works and evaluating effectiveness based on empirical results reported in the surveyed literature. This mapping reveals several critical insights. First, no single countermeasure adequately defends against all threats: gradient inference attacks are effectively mitigated by HE, though this often introduces significant computational latency (e.g., execution time nearly doubling in \cite{mahato2024privacy} or requiring optimized threshold schemes to reduce overhead \cite{jin2025dmafl}); Byzantine poisoning is countered by reputation mechanisms and Deep Reinforcement Learning with accuracy improvements exceeding $80\%$ on MNIST \cite{pan2025privacy}, yet these defenses require complex training procedures and careful threshold tuning; side-channel attacks on TEE implementations are detected with $99.9\%$+ accuracy through SGX-Bouncer monitoring \cite{zhang2021see}, but detection incurs $\times 3.89$ overhead and does not prevent the attack, only detects it post-factum.

Second, defense effectiveness varies dramatically based on adversary capabilities and deployment constraints. CPU monopolization strategies like E-SGX achieve $2.43\%$ overhead when Hyper-Threading is disabled but $47.24\%$ overhead when enabled \cite{lang2020sgx}, demonstrating that seemingly minor platform configuration differences can render defenses impractical. Threshold cryptographic schemes like threshold CKKS \cite{alzahrani5269991privhfl} and threshold TFHE \cite{jin2025dmafl} provide collusion resistance against up to $(n-1)/2$ malicious participants under semi-honest adversarial models, but these guarantees do not extend to malicious adversarial models where participants deviate arbitrarily from protocol specifications.

Third, emerging threats such as rollback attacks and encrypted poisoning attacks are inadequately addressed by existing countermeasures. Rollback attacks against RISC-V TEE implementations remain undefended \cite{bove2023basic}, allowing adversaries to replay old attestation values and potentially bypass security policies. Encrypted model poisoning where adversaries manipulate ciphertexts without knowing underlying plaintexts challenges the assumption that encryption alone provides integrity, necessitating authenticated encryption schemes with additional integrity checks. \rev{Most critically, the integration of Turing-complete smart contracts into decentralized data-sharing frameworks introduces severe availability risks that directly threaten the viability of these systems in real-world deployments. Yaish et al. \cite{yaish2024speculative} demonstrates that adversaries can launch Speculative Denial-of-Service (DoS) attacks, which exploit the transaction verification process to exhaust the computational resources of decentralized nodes comprehensively—all without incurring the expected gas fees. Unlike traditional DoS attacks that require sustained resource expenditure, this speculative variant weaponizes the verification logic itself, forcing nodes to perform expensive computations that are only discarded after the fact. For latency-critical vehicular networks and resource-constrained IoT environments, where continuous data access is not merely a performance metric but a safety-critical requirement, mitigating such economic and resource-exhaustion vulnerabilities is paramount. These attacks reveal that decentralized trust, when implemented naïvely, can become an attack surface rather than a security guarantee—a stark validation of the trilemma's central claim that optimizing for trust alone inevitably compromises availability and efficiency.}

The threat landscape analysis underscores a fundamental principle: effective security requires defense-in-depth through layered countermeasures rather than reliance on any single protection mechanism. Federated Learning must be augmented with both cryptographic protections (HE, Differential Privacy) against inference attacks and algorithmic defenses (reputation systems, Byzantine-robust aggregation) against poisoning attacks. TEE implementations must combine hardware isolation with software-level side-channel mitigations and continuous monitoring to detect exploitation attempts. Blockchain systems must employ both consensus-level Byzantine fault tolerance and application-level access control to prevent unauthorized data manipulation.

\begin{table*}[!t]
\centering
\caption{Mapping Security Threats to Architectural Countermeasures}
\label{tab:threats}
\small
\begin{tabular}{p{3.2cm}p{4.2cm}p{3cm}p{3.5cm}}
\toprule
\textbf{Threat} & \textbf{Countermeasure} & \textbf{Representative Works} & \textbf{Effectiveness} \\
\midrule
\textbf{Gradient Inference Attacks} & Homomorphic Encryption of model updates & \cite{mahato2024privacy,alzahrani5269991privhfl,jin2025dmafl,gadiwala2024enabling,zhao2023tensor} & High privacy but $5.7$-$28.4\times$ overhead \\
\midrule
\textbf{Byzantine Poisoning} & DRL-based weighting, Reputation mechanisms, Clustering & \cite{pan2025privacy,li2023defending,lin2024sf} & Accuracy $+80\%$+ (MNIST); Stable at $50\%$+ attackers \\
\midrule
\textbf{Cache Side-Channels} & CPU monopolization, Load-time synthesis, Runtime monitoring & \cite{lang2020sgx,sang2022pridwen,zhang2021see} & Detection $99.9\%$+; Overhead $2.43\%$-$47.24\%$ (HT-dependent) \\
\midrule
\textbf{Page Table Side-Channels} & T-SGX (TSX), Varys, SGX-Bouncer & \cite{sang2022pridwen,zhang2021see} & Comprehensive mitigation; $2.1$-$3.6\times$ slowdown \\
\midrule
\textbf{Compromised OS} & TEE isolation (TrustZone, SGX), MPU sandboxing & \cite{khurshid2022tee,wang2020trustict,ma2021construction} & Overhead $1.4\%$-$60\%$; Limited secure memory \\
\midrule
\textbf{Key Escrow} & Multi-authority ABE, Threshold schemes, Key escrow removal & \cite{chen2015removing,li2017efficient,alzahrani5269991privhfl} & Eliminates single point; Complexity increases \\
\midrule
\textbf{Data Tampering} & Blockchain immutability, Smart contracts & \cite{shen2022network,jin2025dmafl,ramesh2025privacy,razdan2024decentralized} & Tamper-proof audit; Storage/latency trade-offs \\
\midrule
\textbf{Single Point of Failure} & Decentralized FL (P2P), Multi-server architectures & \cite{li2023prototype,hota2025advanced,karanjai2023dhtee} & Enhanced resilience; High computational cost per device \\
\midrule
\textbf{Data Unavailability (IPFS)} & Pinning, Dynamic replica management, VDICs & \cite{lamichhane2024verifiable,huang2019dynamic} & Verifiable permanency; Write latency increases with nodes \\
\midrule
\textbf{Unauthorized Access (IPFS)} & AES encryption + PRE, VC-based authentication & \cite{mo2024secure,lamichhane2024verifiable} & Fine-grained access control; High gas costs (1.57M Gwei for 10MB) \\
\bottomrule
\end{tabular}
\end{table*}

\subsection{Technology Maturity and Privacy-Utility Trade-off Analysis}
\label{subsec:maturity}

The practical deployment of privacy-preserving architectures in real-world IoT systems is fundamentally constrained by technology maturity and the inherent trade-off between privacy protection strength and system utility (measured by accuracy, latency, throughput, and energy consumption). Table~\ref{tab:maturity} systematically evaluates each architectural paradigm across five dimensions: deployment readiness (Low/Medium/High based on maturity of implementations and existence of production deployments), privacy versus overhead trade-off (qualitative characterization of the balance achieved), suitability for real-time IoT applications (Poor/Fair/Good based on latency requirements), and critical bottleneck (the primary technical barrier preventing widespread adoption).

\textbf{Federated Learning architectures} demonstrate high deployment readiness with multiple production systems deployed by organizations like Google (Gboard keyboard), Apple (Siri), and various healthcare institutions. Vanilla FL without cryptographic enhancements achieves good real-time suitability due to low overhead but suffers from vulnerability to inference and poisoning attacks that can extract sensitive training data or degrade model quality. Integrating HE with FL provides high privacy but introduces $5.7$-$28.4\times$ computational overhead \cite{mahato2024privacy,jin2025dmafl}, reducing deployment readiness to medium and real-time suitability to fair. The critical bottleneck is the computational cost of homomorphic operations which must be performed for every gradient update across potentially thousands of parameters. FL with Differential Privacy achieves a better balance with medium-high privacy and low-medium overhead, though careful noise calibration is required to avoid excessive utility degradation—evidenced by the DP-SA framework achieving $96.38\%$ privacy with $95.12\%$ accuracy \cite{11032530}, demonstrating that thoughtful design can navigate the privacy-accuracy trade-off.

\textbf{Trusted Execution Environment implementations} exhibit medium-high deployment readiness with commercial availability of Intel SGX, ARM TrustZone, and emerging RISC-V solutions. TEEs provide high privacy through hardware-enforced isolation with overhead ranging from $2.43\%$ (E-SGX with HT disabled) to $60\%$ (TrustICT for cryptographic operations) \cite{lang2020sgx,wang2020trustict}, generally maintaining fair-to-good real-time suitability. However, the critical bottleneck is dual: side-channel vulnerabilities that can leak sensitive information despite hardware isolation \cite{zhang2021see}, and limited secure memory (Protected Memory Region in SGX, secure world memory in TrustZone) that constrains the size and complexity of applications that can be protected. The discovery of 70 vulnerabilities through state-aware fuzzing \cite{wang2024syztrust} underscores that TEE security is contingent on correct implementation, not merely hardware features. \rev{This fragility is further exacerbated by fundamental architectural flaws in widely adopted TEE standards. Recent findings presented at \cite{busch2024globalconfusion} revealed that the GlobalPlatform API for ARM TrustZone contains inherent fail-open design weaknesses, systematically exposing Trusted Applications (TAs) to zero-day type-confusion vulnerabilities. Such architectural oversights demonstrate that hardware-enforced isolation can be entirely bypassed through interface exploitation, reinforcing the urgent need for rigorous formal verification of TEE software stacks before deployment in critical IoT and vehicular environments.}

\rev{This prohibitive computational overhead presents a significant barrier for practical deployment. Specifically, most low-power IoT microcontrollers lack the processing capacity and battery life to sustain the $5.7\times$ to $28.4\times$ execution delay imposed by Fully Homomorphic Encryption in real-time vehicular safety scenarios, necessitating more resource-aware hybrid solutions or hardware-assisted isolation \cite{mahato2024privacy, jin2025dmafl, wang2024syztrust}.} FHE provides very high privacy by enabling arbitrary computation on encrypted data without decryption, but even optimized implementations like BFV achieve only 60,000 encrypted image recognitions per hour \cite{peng2023security}—equivalent to approximately 16 images per second for a single CPU, orders of magnitude slower than plaintext inference on modern GPUs achieving thousands of images per second. The critical bottleneck is the computational cost of homomorphic operations which grows with both circuit depth (number of sequential operations) and circuit width (number of parallel operations), exacerbated in IoT devices with limited processing power. Hardware acceleration efforts \cite{mert2022medha,das202310,yang2023poseidon} show promise but have not yet achieved the orders-of-magnitude improvement needed for practical real-time deployment.

\textbf{Partially HE schemes} like Paillier achieve better practical balance, with medium deployment readiness and fair real-time suitability. PHE supports only additive homomorphism, limiting applicability to linear operations but enabling practical implementations for aggregation tasks common in IoT. The critical bottleneck is the limited operation support and the continued need for hardware acceleration to achieve acceptable performance on resource-constrained devices.

\textbf{Attribute-Based and Hierarchical Encryption schemes} demonstrate medium deployment readiness with implementations in enterprise systems requiring fine-grained access control. ABE/HIBE provide high privacy through policy-based encryption but incur medium-high overhead due to expensive bilinear pairing operations that grow with policy complexity. Real-time suitability is fair, suitable for access control decisions but potentially problematic for high-throughput data streams. The critical bottlenecks are expensive pairing operations (often requiring milliseconds per pairing on mobile devices) and key management complexity in hierarchical structures where key derivation paths must be carefully managed to prevent unauthorized access escalation. The privacy-aware VANET authentication scheme achieves practical efficiency through batch verification \cite{zhang2015privacy}, demonstrating that careful protocol design can mitigate computational costs.

\textbf{Blockchain consensus mechanisms} exhibit stark dichotomy between traditional Proof-of-Work (low deployment readiness, poor real-time suitability due to $10+$ minute block times and massive energy consumption) and lightweight alternatives like Proof-of-Authority and Delegated Proof-of-Stake (medium-high deployment readiness, good real-time suitability). The critical trade-off is between decentralization and efficiency: PoW provides maximum decentralization but impractical overhead for IoT; PoA/DPoS achieve acceptable efficiency but centralize trust in validator nodes, potentially creating new single points of failure. Threshold Raft \cite{guo2021location} represents an interesting middle ground, providing Byzantine fault tolerance (tolerating up to $(n-1)/3$ malicious nodes) while avoiding the exponential communication cost increase of classical Raft.

\textbf{IPFS storage solutions} demonstrate high deployment readiness in public form (no privacy, low overhead, good real-time suitability) but face challenges when privacy is required. IPFS with encryption and Proxy Re-Encryption achieves high privacy and medium-high overhead but suffers from scalability issues evidenced by gas costs surging to 1.57M Gwei for 10MB files \cite{mo2024secure}. The critical bottleneck is the economic scalability of blockchain-based access control where transaction costs grow with file size and access frequency, potentially making systems economically infeasible for high-throughput applications. Verifiable Decentralized IPFS Clusters \cite{lamichhane2024verifiable} offer an alternative approach relying on verifiable credentials rather than blockchain transactions for access control, though permanency guarantees depend on voluntary participation by node operators.

\textbf{Hybrid architectures} combining FL, HE, and Blockchain demonstrate medium deployment readiness with emerging implementations showing promising results: $93.2\%$ accuracy with $30\%$ communication reduction in healthcare applications \cite{ramesh2025privacy}, demonstrating that careful integration of complementary techniques can achieve high privacy with medium overhead and fair-to-good real-time suitability. The critical bottleneck is architectural complexity: multi-layer coordination overhead requires sophisticated system design to ensure that security guarantees compose correctly across layers and that failure in one layer does not compromise the entire system. The overhead analysis must account for all layers simultaneously—encryption overhead, consensus latency, and communication costs must be minimized holistically rather than optimized in isolation.

This maturity analysis reveals a fundamental insight: \textit{mature, production-ready solutions tend to occupy narrow regions of the privacy-efficiency-trust design space}, excelling in one or two dimensions while making significant compromises in others. Vanilla FL is efficient and deployable but vulnerable; FHE is secure but impractical; blockchain is trustworthy but slow. Emerging hybrid approaches show promise in expanding the achievable design space by combining complementary techniques, but introduce new challenges in system complexity, composition of security guarantees, and holistic performance optimization.

\begin{table*}[!t]
\centering
\caption{Technology Maturity and Privacy-Utility Trade-off for Real-World IoT Deployment}
\label{tab:maturity}
\small
\resizebox{\textwidth}{!}{
\begin{tabular}{p{3.2cm}p{2.3cm}p{4cm}p{2cm}p{3.5cm}}
\toprule
\textbf{Architecture} & \textbf{Deployment Readiness} & \textbf{Privacy vs.\ Overhead Trade-off} & \textbf{Real-time IoT Suitability} & \textbf{Critical Bottleneck} \\
\midrule
\textbf{FL (Vanilla)} & High & Medium Privacy / Low Overhead & Good & Vulnerable to inference and poisoning attacks \\
\midrule
\textbf{FL + HE} & Medium & High Privacy / High Overhead ($5.7$-$28.4\times$) & Fair & Computational cost; $5.7\times$ time increase \\
\midrule
\textbf{FL + DP} & High & Medium-High Privacy / Low-Medium Overhead & Good & Privacy-accuracy trade-off; Noise degrades utility \\
\midrule
\textbf{TEE (SGX/TrustZone)} & Medium-High & High Privacy / Low-Medium Overhead ($2.43\%$-$60\%$) & Fair-Good & Side-channel vulnerabilities; Limited secure memory \\
\midrule
\textbf{FHE (BFV/CKKS)} & Low & Very High Privacy / Very High Overhead & Poor & Prohibitive computational cost; Impractical for real-time \\
\midrule
\textbf{PHE (Paillier)} & Medium & High Privacy / Medium Overhead & Fair & Limited to additive operations; Hardware acceleration needed \\
\midrule
\textbf{ABE/HIBE} & Medium & High Privacy / Medium-High Overhead & Fair & Expensive pairing operations; Key management complexity \\
\midrule
\textbf{Blockchain (PoW)} & Low & Medium Privacy / Very High Overhead & Poor & Energy consumption; Latency ($10+$ min/block) \\
\midrule
\textbf{Blockchain (PoA/DPoS)} & Medium-High & Medium Privacy / Low-Medium Overhead & Good & Centralized trust in validators; Limited decentralization \\
\midrule
\textbf{IPFS (Public)} & High & No Privacy / Low Overhead & Good & Data publicly accessible; Permanency not guaranteed \\
\midrule
\textbf{IPFS + Encryption + PRE} & Medium & High Privacy / Medium-High Overhead & Fair & High gas costs (1.57M Gwei/10MB); Scalability challenges \\
\midrule
\textbf{Hybrid (FL+HE+BC)} & Medium & High Privacy / Medium Overhead ($30\%$ comm reduction) & Fair-Good & Architectural complexity; Multi-layer coordination overhead \\
\bottomrule
\end{tabular}}
\end{table*}

\subsection{The Privacy-Efficiency-Trust Trilemma: A Fundamental Barrier}
\label{subsec:trilemma}

Our systematic analysis reveals that single-paradigm solutions inherently fail to satisfy the privacy-efficiency-trust trilemma simultaneously. This limitation is not an implementation artifact but a fundamental property of the design space: optimizing any two objectives necessarily compromises the third.

\textbf{Decentralized Computation} (Group 1) excels in computational efficiency through distributed learning and hardware isolation but suffers from adversarial vulnerabilities. Byzantine poisoning attacks degrade FL accuracy from $90\%$ to $21\%$ in worst-case scenarios \cite{lin2024sf}. Gradient inference attacks can reconstruct training samples from shared gradients, violating privacy assumptions. TEE implementations face microarchitectural side-channel attacks achieving $99.9\%$+ detection rates \cite{zhang2021see}, constrained by limited secure memory that fundamentally limits operation complexity.

\textbf{Cryptographic Approaches} (Group 2) provide provable security under well-defined computational hardness assumptions but impose prohibitive overhead. Fully Homomorphic Encryption incurs $5.7$-$28.4\times$ overhead \cite{mahato2024privacy,jin2025dmafl}—even optimized implementations achieve only 60,000 encrypted image recognitions per hour (approximately 16 per second) \cite{peng2023security}, orders of magnitude slower than plaintext inference. ABE/HIBE schemes require expensive bilinear pairing operations that grow with policy complexity, problematic for millisecond-scale vehicular access control decisions.

\textbf{Distributed Ledger Technologies} (Group 3) establish decentralized trust through consensus mechanisms but face scalability bottlenecks. Traditional Proof-of-Work is fundamentally incompatible with resource-constrained IoT devices, requiring massive computational resources ($10+$ minute block times unsuitable for time-sensitive applications). Lightweight consensus mechanisms like PoA/DPoS improve efficiency but centralize trust in validator nodes, creating potential single points of failure. IPFS with encryption and Proxy Re-Encryption incurs gas costs reaching 1.57M Gwei for 10MB files \cite{mo2024secure}, economically infeasible for many applications.

This analysis conclusively demonstrates that the privacy-efficiency-trust trilemma is not merely an implementation challenge but a fundamental property of the design space. The path forward requires architectural integration, combining complementary paradigms rather than incremental optimization within monolithic solutions—a direction we explore in the remainder of this section by identifying critical open challenges and proposing a principled hybrid architectural vision.

\rev{\subsection{Practical Deployment Constraints in IoT and Vehicular Environments}
\label{subsec:practical_constraints}

Before evaluating architectural trade-offs, it is essential to ground the analysis in the \textit{concrete operational constraints} that distinguish IoT and vehicular deployments from general-purpose distributed systems. Our survey explicitly rejects a ``one-size-fits-all'' evaluation, instead assessing each paradigm against the resource budgets, latency requirements, and communication characteristics of real-world deployment targets.

\subsubsection{Vehicular Latency and Mobility Constraints}

V2X communication imposes uniquely demanding requirements that fundamentally shape architectural feasibility. Safety-critical applications such as cooperative collision avoidance and platooning require \textbf{sub-100ms end-to-end latency}---a constraint that renders traditional blockchain mechanisms like Proof-of-Work (with $10+$ minute block times) \textit{fundamentally incompatible} with V2X coordination. Furthermore, vehicular networks exhibit \textbf{highly dynamic validator sets}: vehicles enter and exit communication range within seconds, creating ephemeral trust relationships that static consensus mechanisms cannot accommodate. High-speed handoffs between Road-Side Units (RSUs) introduce additional complexities---authentication and re-keying must complete within the dwell time of a vehicle in RSU range, typically $5$--$30$ seconds at highway speeds. These constraints explain why lightweight consensus mechanisms (PoA, DPoS, Threshold Raft \cite{guo2021location}) and session-bound key management dominate vehicular deployments in our corpus.

\subsubsection{Resource-Constrained IoT Node Profiles}

Our analysis specifically evaluates architectures against the hardware capabilities of \textbf{low-end IoT devices} --  microcontrollers with limited RAM (64--512 KB), constrained processing power (ARM Cortex-M class, typically $\leq$ 200 MHz), and battery-powered operation demanding energy efficiency. This target hardware profile has direct implications:

\begin{itemize}
    \item \textbf{TEE Selection:} ARM TrustZone-M for Cortex-M microcontrollers \cite{khurshid2022tee} and RISC-V Physical Memory Protection (PMP) \cite{bove2023basic} provide lightweight hardware isolation tailored to devices where Intel SGX's memory and power requirements are prohibitive.
    \item \textbf{Cryptographic Feasibility:} FHE's $5.7$--$28.4\times$ overhead \cite{mahato2024privacy,jin2025dmafl} renders it impractical on devices where even plaintext inference is computationally constrained; Partially HE (Paillier) and lightweight symmetric schemes represent practical alternatives.
    \item \textbf{Memory Footprint:} Complex cryptographic libraries and blockchain clients may exceed available RAM on low-end nodes, necessitating split architectures where edge gateways perform heavyweight operations on behalf of constrained devices.
\end{itemize}

\subsubsection{Communication Efficiency as the Dominant Cost}

In wireless IoT deployments, \textbf{communication costs frequently exceed computational costs} by orders of magnitude due to limited bandwidth and high energy consumption of radio transmission. Our analysis identifies several optimizations adopted across the surveyed works:

\begin{itemize}
    \item \textbf{Gradient Compression and Quantization:} Techniques including sparse gradients, top-$k$ selection, and mixed-precision quantization reduce communication volume by up to $70\%$ without significant accuracy degradation.
    \item \textbf{Selective Model Updates:} Layer freezing and partial model sharing strategies transmit only the layers most sensitive to local data distributions, minimizing redundant communication.
    \item \textbf{Hierarchical Aggregation:} Edge-level pre-aggregation in multi-tier FL architectures reduces the number of transmissions to the central aggregator, addressing bandwidth bottlenecks in large-scale deployments \cite{lai2022fedscale}.
\end{itemize}

Table~\ref{tab:deployment_constraints} summarizes the quantitative performance boundaries that define practical feasibility across target deployment environments, derived from empirical results reported in the surveyed literature.

\begin{table*}[!t]
\centering
\caption{Quantitative Performance Boundaries for IoT and Vehicular Deployment Scenarios}
\label{tab:deployment_constraints}
\small
\begin{tabular}{p{3.5cm}p{3cm}p{3cm}p{4.5cm}}
\toprule
\textbf{Constraint} & \textbf{IoT (Low-End)} & \textbf{Vehicular (V2X)} & \textbf{Implication for Architecture Selection} \\
\midrule
End-to-end Latency & $<$ 1s (monitoring) & $<$ 100ms (safety) & FHE infeasible; TEE/DP preferred for real-time \\
\midrule
Computational Overhead & $<$ 10\% acceptable & $<$ 5\% for safety-critical & PHE $>$ FHE; Hardware acceleration required \\
\midrule
Communication Budget & kB-scale per round & MB-scale per round & Gradient compression essential for IoT \\
\midrule
Energy per Transaction & $<$ 1 mJ target & Less constrained (vehicle power) & PoW/PoS infeasible for battery IoT \\
\midrule
Consensus Finality & Seconds & $<$ 1s & PoW ($10+$ min) incompatible; PoA/DPoS viable \\
\midrule
TEE Memory & 64--512 KB (Cortex-M) & MBs (vehicle ECU) & TrustZone-M/RISC-V PMP for IoT; SGX for edge \\
\midrule
Storage Cost (on-chain) & Minimal & Moderate & IPFS+PRE gas costs (1.57M Gwei/10MB) prohibitive \\
\bottomrule
\end{tabular}
\end{table*}

These deployment-specific constraints inform the evaluation criteria applied throughout the remainder of this section, ensuring that our assessment of technology maturity (Table~\ref{tab:maturity}), security trade-offs (Table~\ref{tab:threats}), and hybrid architecture viability reflects the practical realities of target environments rather than abstract theoretical comparisons.}

\subsection{Critical Open Research Challenges}
\label{subsec:challenges}

Our analysis of 75 works identifies four fundamental challenges that must be addressed for practical deployment:

\textbf{Real-Time Performance with Strong Privacy.} The overhead gap between plaintext operations and privacy-preserving variants remains substantial: $5.7\times$ for HE-based FL \cite{mahato2024privacy}, $3.6\times$ for TEE with comprehensive side-channel mitigations \cite{sang2022pridwen}. Achieving $<10\%$ overhead while maintaining semantic security requires hardware-software co-design: specialized accelerators for cryptographic operations \cite{mert2022medha,das202310,yang2023poseidon}, protocols that minimize expensive operations through batching and precomputation, and system architectures that pipeline privacy-preserving operations to hide latency. \rev{In addition to hardware-software co-design, optimizing the execution graph of cryptographic operations is vital. Recent works demonstrate that automated bootstrapping management frameworks, such as DaCapo, can optimally schedule expensive scale-reset operations in Fully Homomorphic Encryption (FHE). This algorithmic optimization significantly reduces the computational bottlenecks of deep FHE multiplication chains, bringing complex privacy-preserving analytics closer to the stringent latency requirements of edge-assisted vehicular networks \cite{cheon2024dacapo}.}

\textbf{Post-Quantum Security for Long-Lived IoT Systems.} Only two surveyed works \cite{hota2025advanced,ali2025privacy} address quantum threats despite IoT systems' decade-long lifecycles. Integration of lattice-based cryptography (NTRU achieving $13\times$ speedup over RSA \cite{hota2025advanced}) into FL aggregation, blockchain consensus, and TEE attestation requires protocol redesign to leverage lattice properties while maintaining efficiency on resource-constrained devices. Critical gaps include efficient threshold lattice cryptography for Byzantine-resilient aggregation and post-quantum consensus mechanisms preserving decentralization.

\textbf{Cross-Platform TEE Interoperability.} Heterogeneous deployments where vehicles, infrastructure, and cloud use different TEE technologies (SGX, TrustZone, RISC-V) lack interoperability frameworks. DHTee's blockchain-based attestation \cite{karanjai2023dhtee} demonstrates feasibility but requires comprehensive performance evaluation and formalization of semantic differences across platforms—SGX provides strong isolation but suffers side-channels; TrustZone has weaker isolation but fewer vulnerabilities; RISC-V implementations vary widely in security properties.

\textbf{Economic Sustainability of Decentralized Systems.} IPFS with blockchain-based access control faces prohibitive costs (1.57M Gwei for 10MB files \cite{mo2024secure}), while voluntary participation models \cite{lamichhane2024verifiable} lack sustainability guarantees. Designing incentive structures that balance economic efficiency, security against economic attacks, and accessibility for resource-constrained devices remains an open problem. Hierarchical approaches where edge devices provide lightweight services rewarded by micropayments while cloud infrastructure handles heavyweight operations show promise but require rigorous game-theoretic analysis.

\subsection{Toward Hybrid Architectures}
\label{subsec:hybrid}

Recent works increasingly adopt multi-paradigm integration in recognition of single-technique limitations. FL combined with HE and Blockchain achieves $93.2\%$ accuracy with $30\%$ communication reduction in healthcare applications \cite{ramesh2025privacy}, demonstrating that careful integration can navigate the trilemma more effectively than monolithic approaches. However, three fundamental challenges persist:

\textbf{Architectural Complexity.} Ensuring security guarantees compose correctly across layers remains difficult. Vulnerabilities in any single layer (insecure FL aggregation despite HE protection, manipulated blockchain consensus despite encrypted storage) can compromise the entire system. The $5.7\times$ time increase in FL with Feldman VSS \cite{shen2022network} illustrates that simply adding cryptographic primitives without holistic co-design yields impractical overhead.

\textbf{Composition of Security Guarantees.} If FL provides differential privacy with parameter $\epsilon_1$ and HE provides computational security with parameter $\lambda$, what are the composed guarantees? Existing analyses reason about components in isolation, but adversaries exploit interfaces between components. Rigorous composition theory for hybrid systems—extending differential privacy composition theorems and universal composability frameworks to multi-paradigm architectures—is needed for end-to-end security guarantees.

\textbf{Multi-Layer Coordination Overhead.} Encryption latency, consensus delay, and communication costs interact in complex ways requiring joint optimization. Practical hybrid systems must minimize end-to-end overhead through techniques including pipelining privacy-preserving operations, adaptive mechanism selection based on threat assessment and resource availability, and hardware acceleration targeting bottleneck operations identified through profiling.

\rev{\textbf{Validation in Heterogeneous Real-World Testbeds.} Last but not least, future research must transition from theoretical simulations to validation in heterogeneous IoT/ vehicular testbeds. Key challenges include:\textbf{(1) Cross-Platform Interoperability:} Developing standardized abstractions that allow seamless data sharing between different hardware TEEs (e.g., SGX and TrustZone) and blockchain protocols \cite{karanjai2023dhtee}; \textbf{(2) Integration Complexity:} Investigating the end-to-end latency when layering multiple primitives (e.g., FL + HE + Blockchain) in high-speed V2X environments where millisecond-scale decisions are critical \cite{adarbah2024new}; \textbf{(3) Energy Sustainability:} Profiling the energy-per-transaction on resource-constrained nodes to ensure that privacy-preserving mechanisms do not lead to premature device failure in large-scale deployments \cite{khurshid2022tee}.}

Addressing these challenges requires interdisciplinary research spanning cryptography (efficient post-quantum protocols, composition theorems), systems (hardware accelerators, cross-platform abstractions), and theory (formal verification frameworks for hybrid systems). The path forward lies not in incremental optimization within individual paradigms but in principled architectural integration informed by systematic analysis of threat models, deployment constraints, and application requirements.


\section{Conclusion}
\label{sec:conclusion}

This survey systematically analyzed 75 privacy-preserving architectures for IoT and vehicular data sharing published between 2007 and 2025, employing PRISMA-guided methodology to classify works into three paradigms: Decentralized Computation (Group 1, 44\%), Cryptography-based approaches (Group 2, 32\%), and Distributed Ledger technologies (Group 3, 24\%). The temporal analysis reveals dramatic acceleration during 2024--2025, with 36 papers (48\% of the corpus) published in this period alone, indicating that privacy-preserving architectures have transitioned from niche research to mainstream concern. Particularly striking is the 2025 surge in Decentralized Computation contributions (10 of 17 papers), establishing distributed learning and hardware isolation as the current research frontier.

Our strategic analysis validates the \textit{privacy-efficiency-trust trilemma}, demonstrating that single-paradigm solutions fundamentally cannot satisfy all three objectives simultaneously. Decentralized Computation (Group 1) achieves efficiency but suffers adversarial vulnerabilities—Byzantine poisoning degrades Federated Learning accuracy from 90\% to 21\%, while Trusted Execution Environments face side-channel attacks and impose 2.43\%--60\% overhead. Cryptographic approaches (Group 2) provide provable security but incur prohibitive costs—Fully Homomorphic Encryption requires 5.7$\times$ to 28.4$\times$ longer execution time, rendering real-time applications impractical on resource-constrained devices. Distributed Ledger technologies (Group 3) establish decentralized trust but face scalability bottlenecks—traditional Proof-of-Work demands 10+ minute block times incompatible with time-sensitive applications, while IPFS with blockchain-based access control incurs gas costs reaching 1.57 million Gwei for 10 megabyte files. This trilemma is not an implementation artifact but a fundamental property of the design space where optimizing any two objectives necessarily compromises the third.

The path forward requires principled architectural integration transcending monolithic approaches. Emerging hybrid systems combining Federated Learning, Homomorphic Encryption, and Blockchain demonstrate promising results—achieving 93.2\% accuracy with 30\% communication reduction in healthcare applications—yet face critical challenges in composition of security guarantees, multi-layer coordination overhead, and cross-platform interoperability. \rev{Our analysis grounds these architectural directions against the concrete operational constraints of IoT and vehicular deployments---including sub-100ms vehicular latency requirements, resource budgets of Cortex-M class microcontrollers, and communication-dominated energy profiles---demonstrating that architecture selection must be deployment-aware rather than technology-driven.} Addressing these challenges demands interdisciplinary research spanning efficient post-quantum protocols resilient to future quantum threats, hardware-software co-design minimizing cryptographic operation overhead to $<10\%$, rigorous composition theory ensuring end-to-end security guarantees across layers, and economically sustainable incentive structures for decentralized systems. The convergence toward multi-paradigm integration observed in recent literature indicates that the research community increasingly recognizes the fundamental limitations of monolithic solutions. Next-generation Intelligent Transportation Systems must embrace hybrid architectures that strategically combine complementary paradigms—leveraging Decentralized Computation for efficiency, Cryptography for provable privacy, and Distributed Ledgers for trust—informed by systematic threat modeling, deployment constraints, and application-specific requirements. \rev{Validating these hybrid architectures in heterogeneous real-world testbeds—addressing cross-platform TEE interoperability, multi-layer integration latency, and energy sustainability on constrained devices—represents a critical next step toward practical deployment.} This survey establishes the principled foundation for hybrid architectural research as the essential trajectory toward practical, deployable privacy-preserving systems meeting the demanding requirements of future distributed intelligence applications.

\section*{Acknowledgement}
We acknowledge the support of time and facilities from Ho Chi Minh City University of Technology (HCMUT), VNU-HCM for this study. 




\bibliographystyle{cas-model2-names}
\bibliography{references}

\end{document}